\documentclass[prd,aps,showpacs,tightenlines,floats,nofootinbib,preprintnumbers]{revtex4}

\usepackage{graphicx}
\usepackage{amssymb}
\usepackage{amsmath}

\newcommand{\xx}{\mbox{\boldmath$x$}}

\newcommand{\unitq}{|\Phi_{\rm in}|^2m}
\newcommand{\unitQ}{|\Phi_{\rm in}|^2}
\newcommand{\unitQQ}{|\Phi_{\rm in}|^2m^{-1}}
\newcommand{\unitEE}{|\Phi_{\rm in}|^2}
\newcommand{\unitQQQ}{|\Phi_{\rm in}|^2m^{-2}}
\newcommand{\unitEEE}{|\Phi_{\rm in}|^2m^{-1}}
\newcommand{\ngrid}{N_{\rm grid}}

\newcommand{\beq}{\begin{equation}}   
\newcommand{\eeq}{\end{equation}}
\newcommand{\bea}{\begin{eqnarray}}   
\newcommand{\eea}{\end{eqnarray}}
\newcommand{\bear}{\begin{array}}  
\newcommand {\eear}{\end{array}}
\newcommand{\bef}{\begin{figure}}  
\newcommand {\eef}{\end{figure}}
\newcommand{\bec}{\begin{center}}  
\newcommand {\eec}{\end{center}}

\def\lrfp#1#2#3{ \left(\frac{#1}{#2} \right)^{#3}}

\begin{document}
%
\title{Numerical study of Q-ball formation in gravity mediation}
%
\author{Takashi Hiramatsu$^1$, Masahiro Kawasaki$^{1,2}$ and Fuminobu Takahashi$^2$}
\affiliation{$^1$Institute for Cosmic Ray Research, The University of
Tokyo, Kashiwa, Chiba 277-8582, Japan}
\affiliation{$^2$Institute for the Physics and Mathematics of the universe, 
  The University of Tokyo, Kashiwa, Chiba 277-8568, Japan}

\begin{abstract}
We study Q-ball formation in the expanding universe on 1D, 2D and 3D
 lattice simulations. We obtain detailed Q-ball charge distributions,
 and  find that the distribution is peaked at
 $Q^{3D}_{\rm peak} \simeq 1.9\times 10^{-2}(|\Phi_{\rm in}|/m)^2$,
 which is greater than the existing result by about 60$\%$.
 Based on the numerical simulations, we discuss how the Q-ball formation
 proceeds. Also we make a comment on possible deviation of the charge
 distributions from what was conjectured in the past.
\end{abstract} 
\maketitle

\preprint{ICRR-Report-565-2009-27}
\preprint{IPMU10-0028}

\section{Introduction}
\label{sec:intro}

If scalar fields are ubiquitous in nature, some of them may remain light in the early universe. 
Such scalar fields could be deviated from the low-energy minimum due to quantum fluctuations during inflation 
and/or modification of the scalar potential through gravitationally suppressed interactions with the inflaton.
At a later time when the expansion rate becomes smaller than the mass, the scalar  will start to roll down
and oscillate about the potential minimum.  The dynamics of such a scalar can exhibit rich phenomena in cosmology.

We focus on a complex scalar field with a global U(1) symmetry at the potential minimum. 
The U(1) symmetry is not necessarily exact, and could be violated at high energy scales; our arguments apply if it 
is a good symmetry at low energy. If supersymmetry (SUSY) is realized in nature, there may be many such
 scalar fields. In Ref.~\cite{Coleman:1985ki} Coleman found that, if the scalar potential is shallower than a quadratic potential, there is a non-trivial field 
 configuration that minimizes the energy for a fixed U(1) charge $Q$. Since the solution has a
 spherical symmetry, the solution is named Q-ball.
A Q-ball is a non-topological soliton whose stability is
supported by the U(1) symmetry.

 It was noticed in Refs.~\cite{Kusenko:1997zq,Dvali:1997qv,Kusenko:1997si,Enqvist:1997si} that Q-balls play an important role
in a context of the Affleck-Dine (AD) mechanism~\cite{Affleck:1984fy}.
The mechanism utilizes a flat direction of the supersymmetric standard model (SSM), which possesses a
non-zero baryon (or lepton) number.  A flat direction responsible for
the AD mechanism is referred to as the AD field (denoted by $\Phi$ in the following). 
The AD field $\Phi$  develops a large expectation value during inflation, and it
starts to oscillate after inflation when the cosmic expansion rate
becomes comparable to its mass. The baryon number is effectively
created at the onset of the oscillations. Finally, $\Phi$ decays
into the ordinary quarks, leaving the universe with a right amount of
the baryon asymmetry. It was realized however that, soon after the onset of oscillations,
the AD field experiences spatial instabilities and deforms into clumpy
Q-balls~\cite{Kusenko:1997si}. Later it was shown in \cite{Kasuya:2000wx}
that most of the baryon asymmetry is absorbed into Q-balls.
Its non-linear property necessitates numerical approach to the formation
and the subsequent evolution. 

The properties of Q-balls crucially depend on the size of the U(1) charge $Q$.
For instance, the mass per unit charge is a decreasing function of $Q$ in case of gauge-mediation type Q-balls;
for a large enough $Q$, the Q-ball can be absolutely stable against the decay into nucleons 
and therefore contribute to dark matter~\cite{Kusenko:1997si}.
In the case of unstable Q-balls, the decay rate depends on the charge, and Q-balls can be very long-lived in a cosmological time scale
for a sufficiently large $Q$~\cite{Enqvist:1998xd},  since the Q-ball decay process takes place only in the vicinity of the surface~\cite{Cohen:1986ct}.
Thus it is of utmost importance to determine the charge distribution of the Q-balls at the formation. 

There is a number of numerical studies on the Q-balls.
The Q-ball formation has been first examined by Kasuya and one of the authors (MK) 
on 3D lattice simulations in a context of gauge mediation~\cite{Kasuya:1999wu,Kasuya:2001hg} and 
gravity mediation~\cite{Kasuya:2000wx}, taking account of the cosmic expansion. They found that most of the baryon 
number is absorbed into the largest Q-balls in the lattices and
estimated the typical Q-ball charge.
However, the estimated charge might have been affected by relatively coarse spatial resolution due to the limited computational power
at that time. The Q-ball charge distribution was studied in 2D~\cite{Enqvist:2000cq} and 3D~\cite{Multamaki:2002hv} lattices,
with an emphasis on negative Q-balls formed for an orbit with a large ellipticity. In gravity mediation, however,
the AD field naturally acquires a large kick into the phase space, leading to an orbit with a comparatively small ellipticity.
As we will see later, the formation of negative Q-balls are also observed in our simulations for a large ellipticity. However,
the final charge distribution does not fit well the empirical formula conjectured in Ref.~\cite{Enqvist:2000cq} based on the entropy argument.
The difference may be attributed to the fact that the charge re-distribution through exchanges of small secondary Q-balls seems less efficient
in our simulations.
Also the finite grid size could have affected the Q-ball evolution especially at a very late time when one Q-ball was represented
by a single grid point. In our analysis we stop following the evolution once the spatial resolution becomes insufficient to describe the
Q-ball solution, in order to avoid any artificial effects due to the
finite grid size. Recently Tsumagari studied the Q-ball formation and
thermalization processes in a great detail in a Minkowski space
neglecting the cosmic expansion \cite{Tsumagari:2009na}. In an expanding
universe, however, the thermalization processes through Q-ball
interactions may be weakened,  and the resultant Q-ball configuration
could be quite different. We will see in the following that the cosmic
expansion indeed plays an essential role in  the Q-ball formation.

In this paper we study the Q-ball formation in
a scalar potential motivated by the AD mechanism in gravity mediation
and obtain the Q-ball charge distribution on 1D, 2D and 3D lattice simulations with the cosmic expansion
for different values of ellipticity and strength of instabilities. We develop a sophisticated numerical code based on
the 6th-order symplectic integrator equipped with a new Q-ball identification algorithm. 

The paper is organized as follows. In Sec.~\ref{sec:dynamics}, we
summarize the governing equation of Q-balls in the gravity-mediation
model. In Sec.~\ref{sec:setup}, we briefly mention the numerical
scheme used in the simulations and, in Sec.~\ref{sec:numerical}, we show
numerical results obtained from simulations in 1D, 2D and 3D configurations.
After that, Sec.~\ref{sec:applications} is devoted to some applications
of our numerical results, and we conclude in Sec.~\ref{sec:conclusion}.

\section{Q-ball dynamics}
\label{sec:dynamics}

In the gravity-mediated SUSY breaking model, the AD field typically 
has a potential of the following form,
%
\begin{equation}
  V(\Phi) = m^2|\Phi|^2 \left[1+K\log\left(\frac{|\Phi|^2}{M_*^2}\right)
			\right]-cH^2|\Phi|^2 + \frac{\lambda^2}{M^{2n-6}}|\Phi|^{2n-2},
\label{eq:ADpotential}
\end{equation}
%
where $\Phi$ is a complex scalar field parametrizing the flat direction,
$\lambda$ a coupling constant of the non-renormalizable term,
$H$ the Hubble parameter, $c$ a positive constant, $M$ a cutoff scale of
the theory, $M_*$ the renormalization point defining the mass of scalar
field $m$, and $K=-0.01 \sim  -0.1$ is the contribution from the one-loop
correction chiefly due to gauginos~\cite{Enqvist:1997si}. 
The cut-off scale $M$ is typically taken as
the GUT scale $\sim 10^{16}$ GeV 
or the Planck scale. Note that the potential (\ref{eq:ADpotential}) is invariant
under the phase rotation of $\Phi$. The effect of the U(1)-violating term will
be taken into account in the initial condition of $\Phi$.

The evolution equation of the scalar field $\Phi$ is given by 
$\nabla_\mu \nabla^{\mu}\Phi= -dV(\Phi)/d\Phi^\dag$.
With the above potential, the explicit form of the governing equation of
$\Phi$ becomes
%
\begin{equation}
\ddot{\Phi} + DH\dot{\Phi}-\frac{1}{a^2}\nabla^2\Phi + m^2\Phi
  \left[1+K+K\log\left(\frac{|\Phi|^2}{M_*^2}\right)\right]
  -cH^2\Phi + \frac{(n-1) \lambda^2}{M^{2n-6}}|\Phi|^{2n-4}\Phi = 0,
\label{eq:ADevo}
\end{equation}
%
where $D$ is the spatial dimension, the dot indicates a time derivative,
and $\nabla^2$ is defined as the Laplacian with respect to the comoving
coordinate. 
In this paper, we restrict our analysis to the epoch between the end of 
inflation and the reheating era, where the scale factor behaves as
$a\propto t^{2/3}$ like the matter-dominant case.
In addition, the non-renormalizable term (the last term being proportional to
$\lambda$) does not directly contribute to the late-time evolution 
of the scalar field, so we fix $\lambda=0$, and, for simplicity, $c=1$. 

The charge density of the AD field is given by
%
\begin{equation}
  q(t,\xx) = -i(\Phi^*\dot{\Phi}-\Phi\dot{\Phi}^*).
\label{eq:q}
\end{equation}
%
The charge and energy of a Q-ball are given by
%
\begin{align}
  Q &= \int_V q(\xx)\, d^Dx ,
\label{eq:Q} \\
  E &= \int_V \left[ |\dot{\Phi}|^2 + |\nabla\Phi|^2 + V(\Phi) \right]\, d^Dx ,
\label{eq:E}
\end{align}
%
where $V$ denotes the $D$-dimensional volume of the Q-ball.
Note that, from the dimensional analysis, these quantities in 1D and 2D
should be regarded as the densities per unit area and length, respectively.

For a negative value of $K$, the potential (\ref{eq:ADpotential}) is shallower than the quadratic potential $m^2|\Phi|^2$, 
for which there exists a non-trivial field configuration that minimizes the energy for a fixed charge $Q$. 
We can find the solution using the Lagrange multiplier method, and the minimum energy state turns out to be a spherical
field configuration inside which the scalar field is rotating in a circular orbit in the phase space~\cite{Coleman:1985ki}. 
This object is called Q-ball.

There is an instability band for the potential with $K<0$. From a simple linear analysis~\cite{Kasuya:2000wx} one can show that the instability
band is given by \footnote{In Ref.\cite{Kasuya:2000wx}, there is
an error in the instability band. One of the authors has rederived the
instability band with a more plausible assumption in
Ref.\cite{Enqvist:2002si}.} 
%
\begin{equation}
  0 < \frac{k^2}{a^2} < 2m^2|K|.
\label{eq:inst}
\end{equation}
%
The small fluctuations $\delta \Phi$ within the above instability band will
therefore grow into Q-balls with a typical size 
$R\sim |K|^{-1/2}m^{-1}$. Since the Q-ball formation is a highly
nonlinear phenomenon, we need to rely on numerical simulations to
understand how it proceeds. 

\section{Simulation setup}
\label{sec:setup}

The potential (\ref{eq:ADpotential}) is invariant under the U(1)
rotation of $\Phi$, and therefore the charge $Q$ is preserved.  In
general, however, there is a U(1)-violating term arising from the A-term
$\propto (\Phi^n + \Phi^{*n})$. This term kicks the AD field into the
angular direction at the onset of oscillations, giving a nonzero U(1)
charge, or equivalently, the angular momentum in the complex $\Phi$
plane [see Eq.~(\ref{eq:q})]. Since the amplitude of the oscillations
decreases due to the Hubble friction term, such non-renormalizable
U(1)-violating terms soon become negligible. Thus, the main role of the
A-term is to give an initial angular momentum. To take account of the
effect, we parametrize the initial condition of the homogeneous mode as 
%
\begin{equation}
 \Phi(t_{\rm in})=M_*,\quad \dot{\Phi}(t_{\rm in})=i\epsilon M_*m.
\label{eq:init}
\end{equation}
%
The initial charge density is then
$q(t_{\rm in})\simeq 2\epsilon M_*^2m$. 
In the following analysis we take $\epsilon = 0.01,0.1,$ and $1$.
Also we add small fluctuations to the above initial condition, 
$|\delta\Phi/\Phi| =10^{-7}$, following Ref.~\cite{Kasuya:2000wx}.

We use a comoving box as a computational domain whose physical size
at the initial time is $a(t_{\rm in})L=bH_{\rm in}^{-1}$,
where $H_{\rm in}$ is the Hubble parameter at the initial time, and $b$ is a
dimension-less parameter defining the box size. We choose the
normalization of the scale factor as $a(t_{\rm in})=1$ where 
$t_{\rm in}$ is the initial time, $t_{\rm in}=2/(3H_{\rm in})$. 
We adopt a value between $b=1\sim 5$. 
In all simulations except for the partial cases (1D2, 1D3, 1D4 in Table \ref{tab:1D}) in 1D, we use the equation of motion in an expanding background
with an initial condition $H_{\rm in}=m$.  In the non-expansion case,
$L$ coincides with the physical scale.  
As for the computational grid, we use the homogeneous $\ngrid^D$ 
grid, where $D$ is the dimension and 
$256\leq \ngrid \leq 4096$ for $D=1$, $\ngrid=512$ for $D=2$ and
$\ngrid=128$ for $D=3$. 
Other numerical parameters we used are tabulated in Tables
\ref{tab:1D}-\ref{tab:3D}.
On the boundaries, we impose the periodic boundary condition.

As the universe expands,
the spatial resolution gets worse. The physical Q-ball size is given
by $R\sim |K|^{-1/2}m^{-1}$~\cite{Enqvist:1997si}, and the grid spacing is
$\Delta x = a(t)L/\ngrid$. Requiring $R > \Delta x$ during the
simulations, 
we obtain the critical time after which Q-balls are no longer resolved,
%
\begin{equation}
 \tau_{\rm c} \equiv m(t_{\rm c}-t_{\rm in})  \simeq
\frac{2}{3}\left(\frac{\ngrid}{b|K|^{1/2}}\right)^{3/2},
\label{eq:tcrit}
\end{equation}
%
where $\tau$ measures the time normalized by $m^{-1}$ from the beginning
of the simulation, i.e., $\tau \equiv m(t-t_{\rm in})$. For example,  we
can resolve Q-balls until $\tau < 1.1\times 10^4$ for
$\ngrid=1024$ and
$b=5$. In all simulations we have chosen the number of grids so that
Q-balls are resolved until $\tau \sim 10^4$.

To evolve $\Phi$, we implement the 7 stages 6th-order symplectic
integrator developed by Yoshida \cite{Yoshida:1990zz}. This scheme
conserves the charge and energy of the target field, and the 
higher-order scheme is expected to suppress the global error, at least,
until the epoch of the Q-ball formation. 
The spatial derivatives, i.e. the Laplacian operator in Eq.~(\ref{eq:ADevo}),
are estimated with the second-order finite difference.
We also implemented the Fourier collocation method, and confirmed
that the final results do not depend on the scheme of derivative.

In order to identify the formed Q-balls, we developed an algorithm based on the
Marching Cube method \cite{Lorensen:1987}. We briefly summarized the
algorithm in 
Appendix B. Our algorithm regards the region in which the charge
density exceeds a critical value, $|q(t, \xx)| > q_c$, as a
Q-ball. This criterion is determined so that $q_c$ is sufficiently
larger than the background field but smaller than the typical peak density.
We set $q_c$ to be a fraction of $q(t)$ when the fluctuations first enter the non-linear regime.

In addition, we use OpenDX to visualize the Q-balls in the 2D and 3D
cases \cite{opendx}.

\section{Numerical results}
\label{sec:numerical}

\subsection{1D}
\label{subsec:1D}

\subsubsection{the effect of cosmic expansion}
\label{subsubsec:exp1D}

Let us begin with the time-evolution of the AD field in 1D. For the sake
of seeing the effect of the cosmic expansion particularly in 1D, we have
performed simulations both with and without the cosmic expansion.

\begin{figure}[!ht]
\centering{
\includegraphics[bb=0 0 297 297, width=8cm]{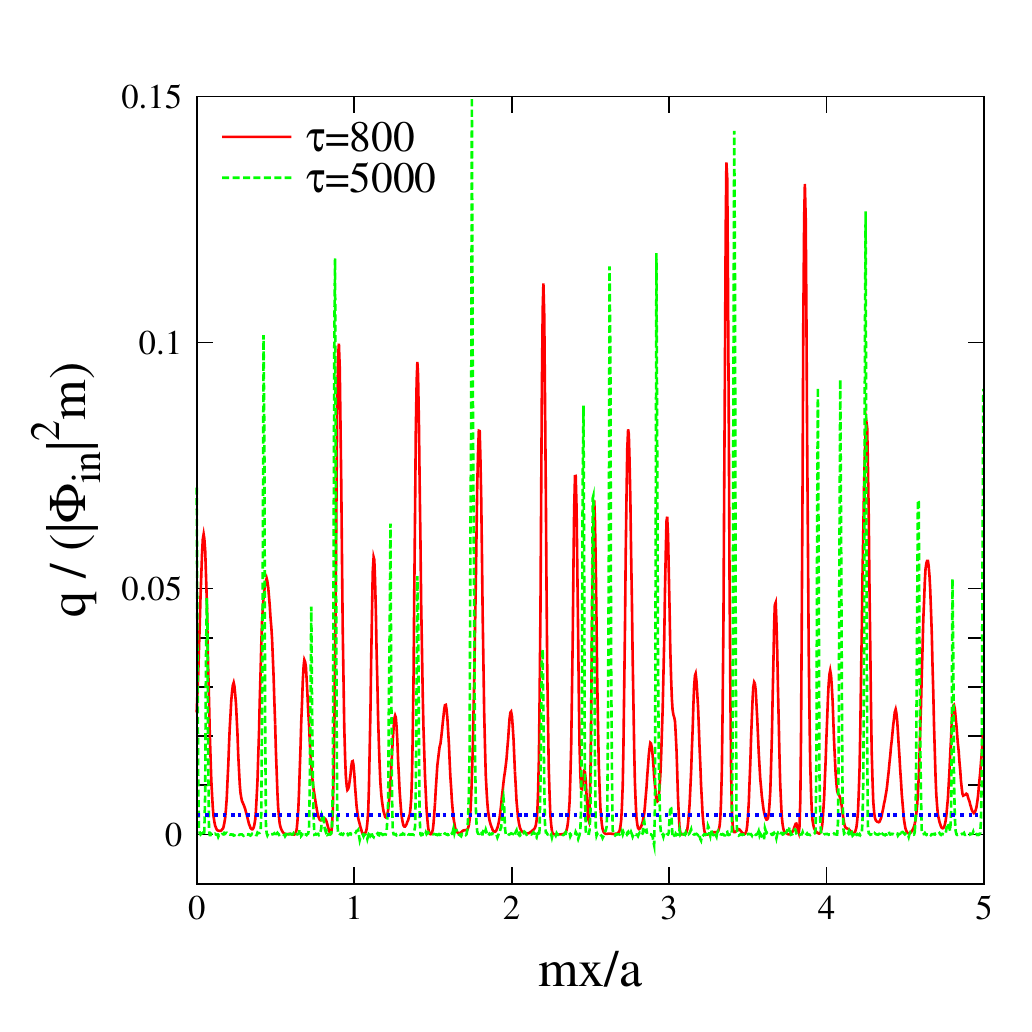}
}
\caption{The spatial charge-density distribution in the simulation 1D1. The blue dotted
 line is the criterion we set for this simulation, given by
 $q_{\rm c}=4\times 10^{-3}\unitq$. The horizontal axis is the comoving
 coordinate.
} 
 \label{fig:field1D}
\end{figure}

First, we show the result with the cosmic expansion and explain how to
identify Q-balls. Fig.~\ref{fig:field1D} shows the charge density of the
AD field at $\tau=800$ and $5000$ with the parameter set 1D1 (see Table
\ref{tab:1D}). There are lots of peaks to be regarded as Q-balls. These
Q-balls were formed at $\tau\sim 700$ in this simulation when the
background charge density has been decreased to $q_{\rm form} \sim
q_0/a(700) \sim 0.02\, \unitq$ where $\Phi_{\rm in}$ is the initial
value of $\Phi$ given in Eq.~(\ref{eq:init}). Once a Q-ball is formed,
its charge density at the center remains virtually unchanged unless
there are significant  collisions and mergers, since its physical size
does not change in time and the total charge is preserved under the
theory we consider here. Actually, we can see that the peak values are
more or less unchanged between $\tau=800$ (red solid) and $\tau=5000$
(green dashed) in Fig.~\ref{fig:field1D}. 

At the formation, the modes in the instability band (\ref{eq:inst}) grow
and become non-linear, and proto-Q-balls are formed. Those proto-Q-balls
absorb the surrounding charges, and the charge density outside Q-balls
are quickly damped. Therefore if we set $q_c$ to be a fraction of
$q_{\rm form}$, we can collect the  Q-balls. Throughout this paper, we
choose $q_{\rm c} = q_{\rm form}/5$. The explicit value of $q_c$ depends
on each simulation. In the present case, 
$q_{\rm c}=4\times 10^{-3} \unitq$, which is drawn as the blue dotted
line in Fig.~\ref{fig:field1D}.

\begin{figure}[!ht]
\centering{
\includegraphics[bb=0 0 311 311, width=8cm]{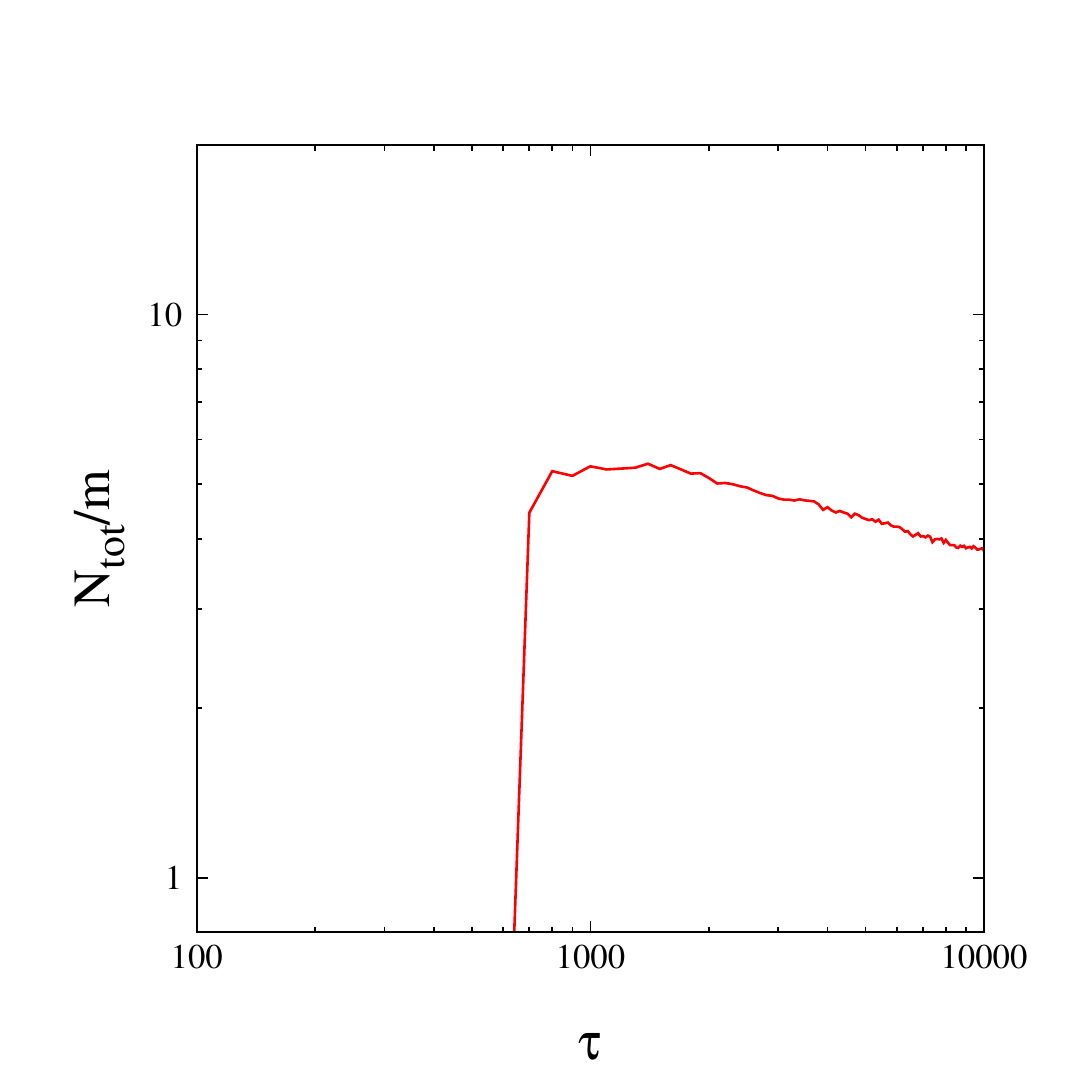}
}
\caption{The time evolution of the total number of Q-balls in the unit
 comoving length,  
for the parameter set 1D1 and $q_{\rm c}=4\times 10^{-3}\unitq$. 
} 
 \label{fig:ncount1D}
\end{figure}

Using this criterion, we show the time evolution of the number of
Q-balls in Fig.~\ref{fig:ncount1D}. The vertical axis is the number of
Q-balls in unit (comoving) length of the computational domain,
$N_{\rm tot}$. We found that there are approximately 5 Q-balls in the
unit comoving length 
(normalized by $m$), that is, totally $\sim 25$ Q-balls in the whole
domain with $b=5$. To reduce the statistical error, we averaged 50
realizations. We can see that, after Q-balls are formed at about $\tau
\sim 700$, the number of Q-balls remains constant for a while and then
gradually decreases. The decrease is probably due to the merger or
disruption process through collisions.  In the 1D simulations,
collisions among Q-balls are expected to occur more frequently than
in the cases with 2D or 3D, which enhances the probability that multiple
Q-balls merge into a large one. Note that this is not due to the finite
box effect, as will become clear from the following argument. 

\begin{table}[!ht]
\begin{tabular}{c|cccccc}
\hline
ID & $\#$ of grid & $Lm$ & $dt$ & $K$ & expansion & realization \\
\hline\hline
1D1 & 1024 & 5 & 0.005 & -0.1 & yes & 50 \\
1D2 & 256 & 100 & 0.05 & -0.1 & no & 50 \\
1D3 & 1024 & 400 & 0.05 & -0.1 & no & 50 \\
1D4 & 4096 & 1600 & 0.05 & -0.1 & no & 20 \\
\hline
\end{tabular}
\caption{Numerical parameters for 1D simulations.}
\label{tab:1D}
\end{table}

\begin{figure}[!ht]
\centering{
\includegraphics[bb=0 0 283 283,width=8cm]{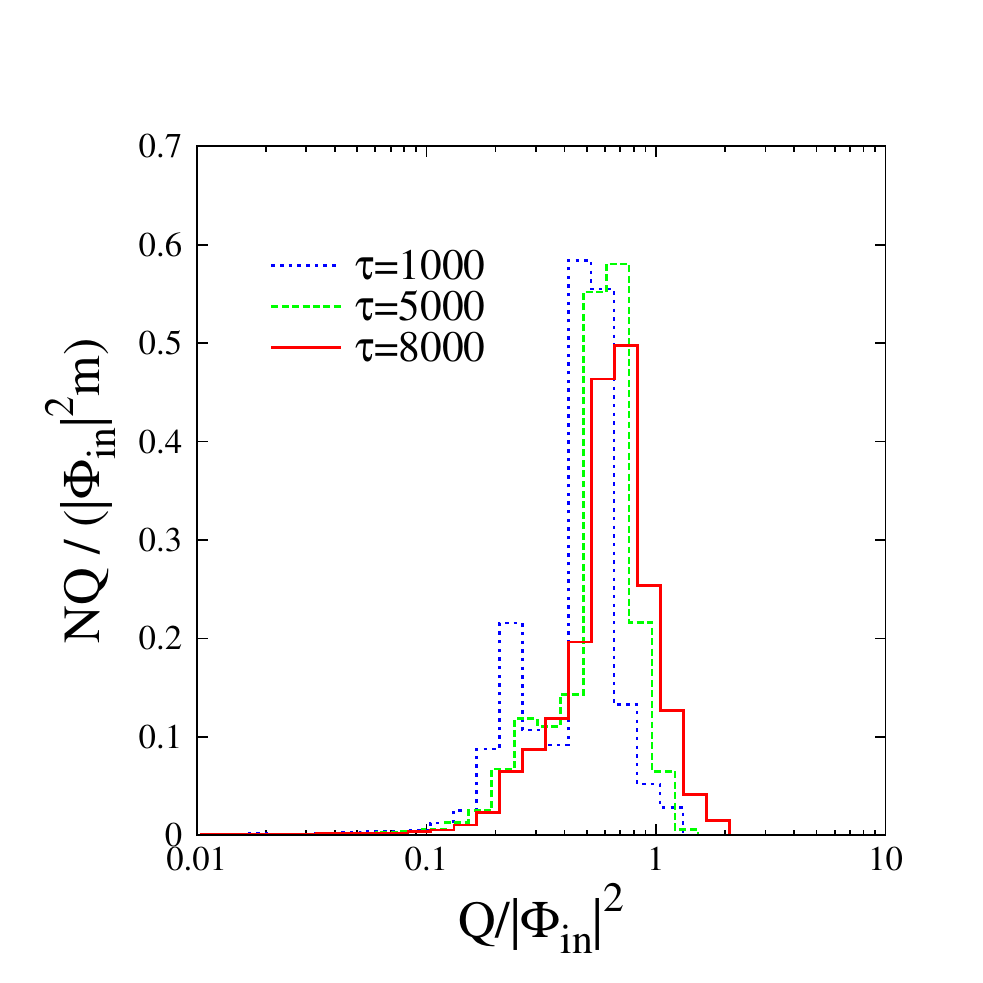}
}
\caption{The charge distributions of Q-balls at $\tau=1000$ (blue
 dotted), $\tau=5000$ (green dashed) and $\tau=8000$ (red solid), for
 the parameter set 1D1 and  $q_{\rm c}=4\times 10^{-3}\unitq$. The
 vertical axis is the number of Q-balls with charge $Q$
 in a comoving length, $N(Q)$, multiplied by the charge. 
} 
 \label{fig:charge1D}
\end{figure}

Identifying each Q-ball, we can estimate its charge and energy. The
charge distribution of Q-balls at $\tau=1000$ (blue dotted), $\tau=5000$
(green dashed) and $\tau=8000$ (red solid) is shown in
Fig.~\ref{fig:charge1D}. The vertical axis is the number multiplied by
the charge. Note that the number here means the one in a logarithmic
bin in a unit comoving length, $N(Q)$. The charge distribution does not 
significantly change in time 
except for a temporal bimodal feature at $\tau = 1000$, and the Q-balls
mainly contributing to the total charge in the computational domain are
those with $Q\simeq \unitQ$. Note that $Q$ should be interpreted as the
charge density per unit area, and so, it has a mass dimension two.
Moreover we have found that the Q-ball charge ranges over about one
order of magnitude, $0.2 \lesssim Q/\unitQ \lesssim 2$. 

In Ref.~\cite{Kasuya:2000wx}, it was concluded that based on the 3D
simulation, a small number of very
large Q-balls absorb most of the charge that the AD field originally
possessed. We however have found that the charge has a distribution
with a finite width 
shown in Fig.~\ref{fig:charge1D}. Therefore, in this paper, we regard  the
typical charge of Q-balls as that at the peak of the charge distribution
$NQ$ instead of the rare largest Q-balls.

So far, we consider the case with the cosmic expansion. In the
non-expansion case, the spatial resolution does not change during
simulations. So, it may sound good for the purpose of numerical
simulation and there are some past works assuming the non-expanding
background, e.g. \cite{Tsumagari:2009na}. However, there are several
crucial differences. 

\begin{figure}[!ht]
\centering{
\includegraphics[bb=0 0 311 311,width=8cm]{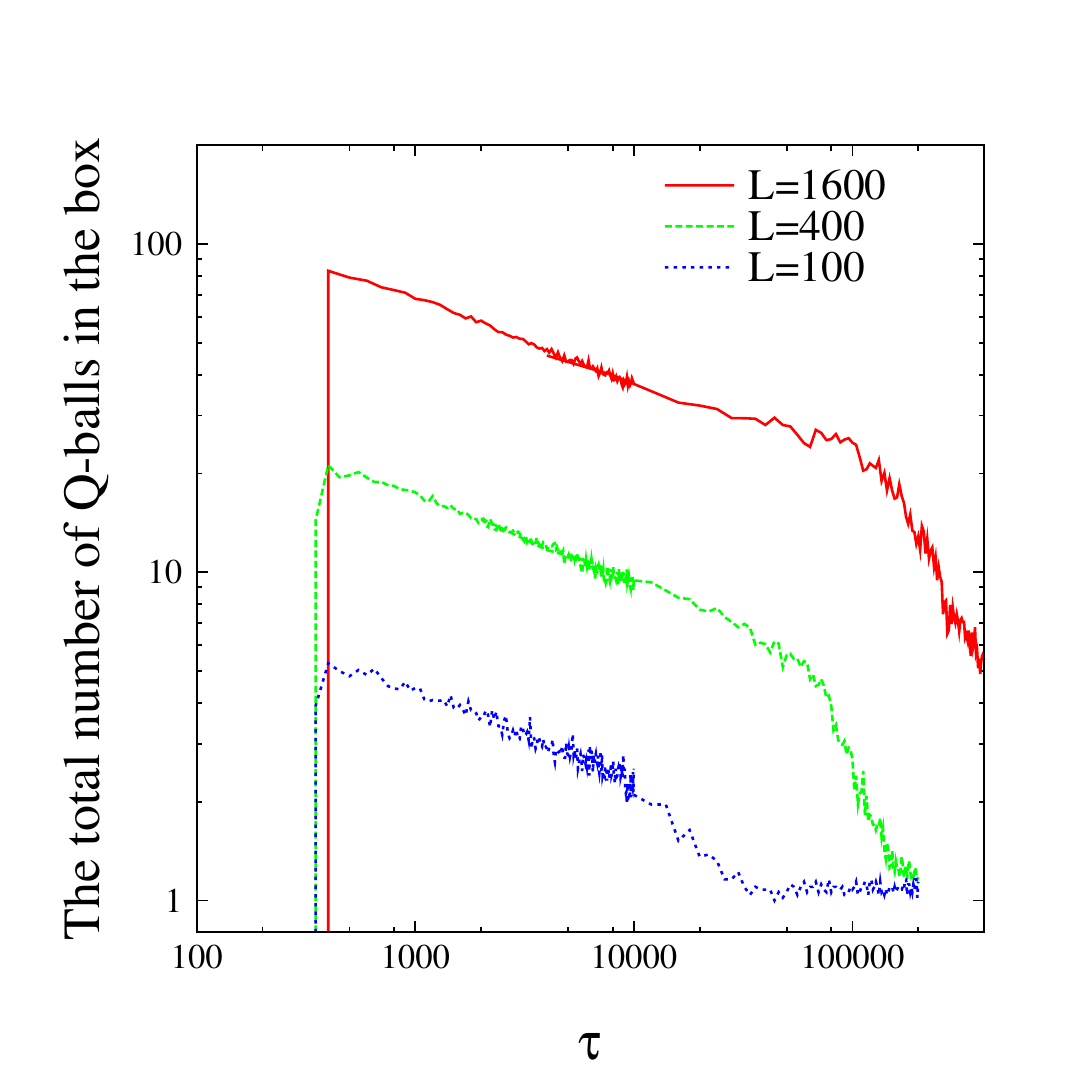}
}
\caption{The evolution of the total number of Q-balls in
 the unit comoving length in non-expansion
 cases. The box size $Lm$ are $100, 400$ and $1600$ while the spatial
 resolution of these simulations are fixed.
 For $\tau >10^4$, we expand the output interval to avoid the
 dense plot.
} 
 \label{fig:ncount1D_non}
\end{figure}

Fig.~\ref{fig:ncount1D_non} shows the time evolution of the number of Q-balls in
the non-expansion cases with the parameters 1D2, 1D3 and 1D4 tabulated
in Table \ref{tab:1D}. The red solid, green dashed and blue dotted lines
are the results with a different box size: $Lm=100, 400$ and $1600$,
respectively. The vertical axis is the total number of Q-balls in a
simulation box. These simulations have the same spatial resolution 
$\Delta x=Lm/\ngrid$. In order to reduce the statistical fluctuations, we
average over 50 realizations for $Lm=100, 400$ and 20 realizations for
$Lm=1600$.  The Q-balls are formed at $\tau \sim 400$, and their number
monotonically decreases due to merger and disruption through
collisions. The decrease may be similar to what we observed in the case
with the cosmic expansion. However, at about 
$\tau \sim 2\times10^4, 7\times 10^4, 1\times 10^5$ for
$Lm=100, 400, 1600$, respectively, the Q-ball number starts to rapidly
decay. In fact, the numbers tend to converge to 1, and the convergence
time as well as the time when the rapid decay starts get delayed for a
larger box. This observation shows  that the large Q-balls sweep the
whole computational domain many times, absorbing other small Q-balls,
until one very large Q-ball is formed.  This is clearly the artificial
volume effect.  Therefore the numerical results in the non-expansion
cases can be trusted only before the time when the numbers start to
rapidly decay. 

\begin{figure}[!ht]
\centering{
  \includegraphics[bb=0 0 793 566,width=12cm]{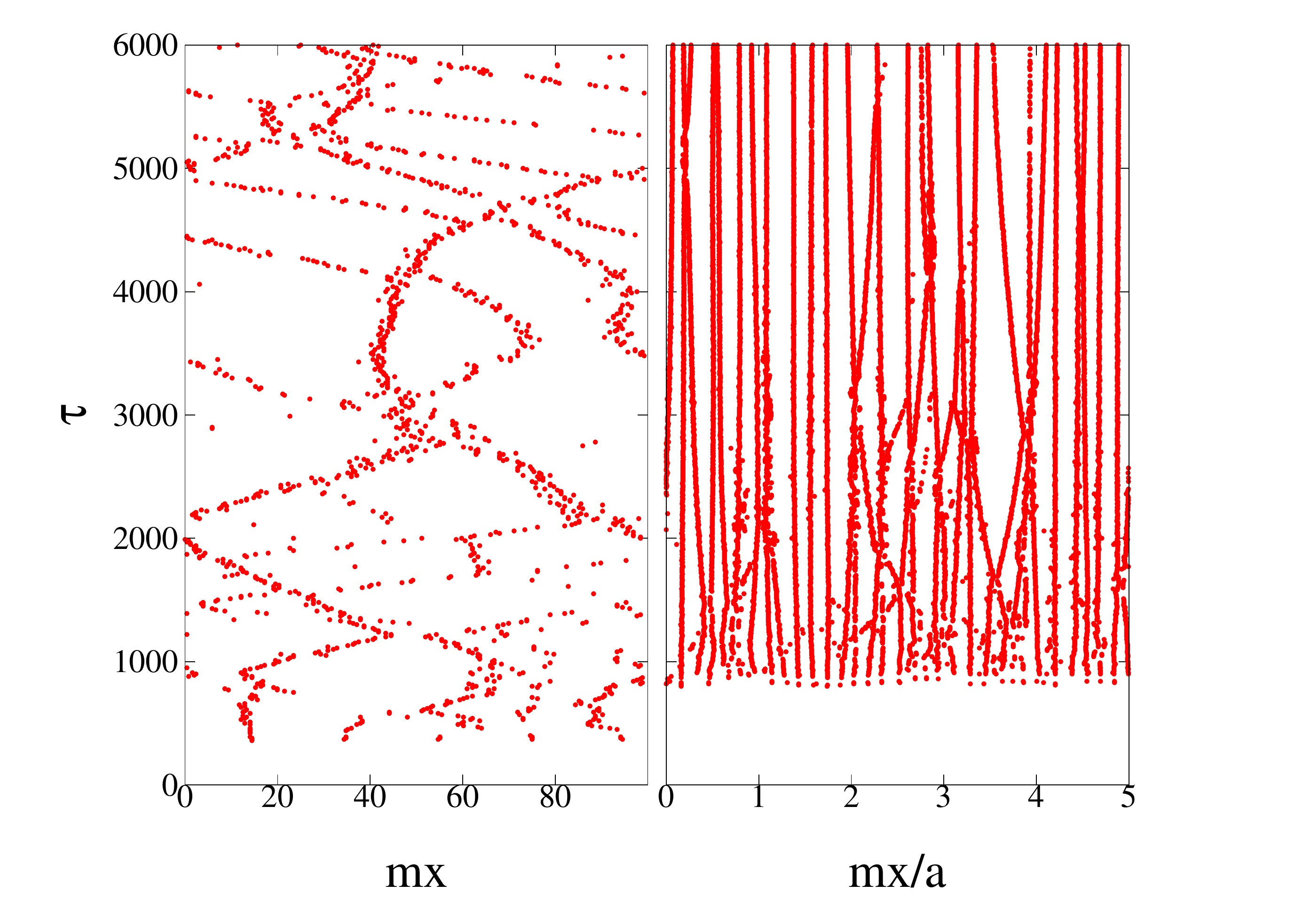}
}
\caption{Trajectories of the Q-balls in the cases without ({\it left}) and with  ({\it right})
the cosmic expansion. Each dot represents the
 position of a Q-ball. The horizontal axis of the
 right panel is the comoving coordinate, and that of the left one is 
 the physical coordinate in unit of $m^{-1}$.
} 
 \label{fig:trajectory}
\end{figure}

The difference between the non-expansion and expansion cases can be
manifestly seen from Fig.~\ref{fig:trajectory}. This figure shows the
trajectories of Q-balls in the case with the cosmic expansion (1D1 in
Table \ref{tab:1D}) [{\it right}] and without (1D3) [{\it left}]. In the
non-expansion case, we found that the Q-balls interact many times with
one another, and Q-balls go across the boundaries of the domain again
and again. Through the interactions among  Q-balls, large Q-balls tend
to become even larger by ingesting other slow-moving Q-balls, resulting
in the decrease of the number of Q-balls. Since the background spacetime
is static, the process will never stop until all the charge is absorbed
into one large Q-ball. Hence, in order to avoid the artificial volume
effect, a large computational box must be prepared. 

In the expanding background, Q-balls are  dragged by the cosmic
expansion and thus less frequently interact with other Q-balls. In fact,
as shown in the right panel of Fig.~\ref{fig:trajectory}, the number
does not change so much even with a quite small box. The cosmic
expansion clearly suppresses the interactions among Q-balls, and the
merger and disruption processes do not frequently take place; they are
considered to be decoupled at a certain point.  In the 2D and 3D cases,
the finite box effect may look weaker than in the case of 1D, since the
collisions are expected to take place less frequently. However, the
intrinsic problem of neglecting the cosmic expansion mentioned above
still exists even in these cases, and it is quite likely that such a
crude approximation significantly affects the Q-ball formation and the
final Q-ball charge distribution. Thus we hereafter will treat only the
cases with the cosmic expansion.

\subsection{2D}
\label{subsec:2D}

\subsubsection{Filamentary structure}
\label{subsubsec:structure2D}

\begin{figure}[!ht]
\begin{tabular}{cc}
\begin{minipage}{6cm}
\includegraphics[bb=0 0 503 478,width=6cm]{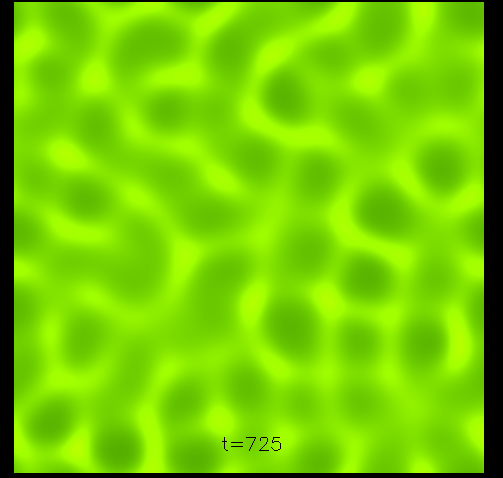}
(a) $\tau=725$
\end{minipage}
\begin{minipage}{6cm}
\includegraphics[bb=0 0 503 478,width=6cm]{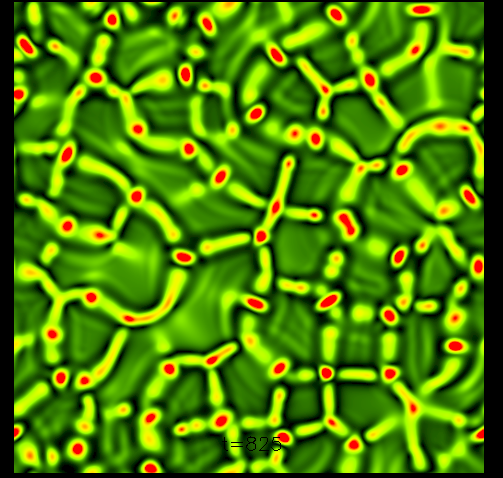}
(b) $\tau=825$
\end{minipage}
\\
\begin{minipage}{6cm}
\includegraphics[bb=0 0 503 478,width=6cm]{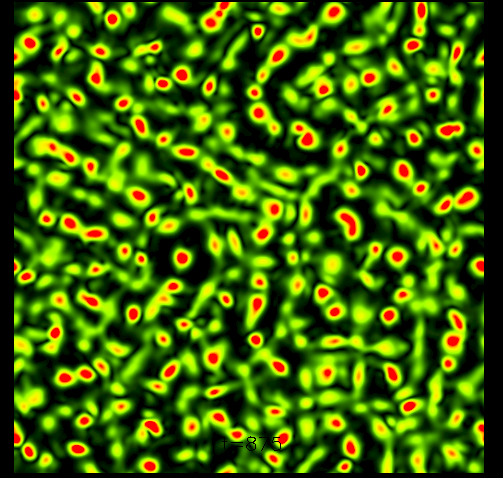}
(c) $\tau=875$
\end{minipage}
\begin{minipage}{6cm}
\includegraphics[bb=0 0 503 478,width=6cm]{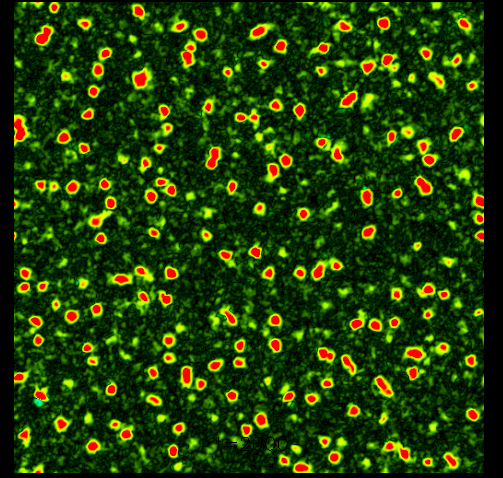}
(d) $\tau=2000$
\end{minipage}
\end{tabular}
\caption{Time evolution of charge density of the AD field in 2D. The red
 indicates the large charge region. For these visualizations, we used OpenDX.
} 
 \label{fig:structure2D}
\end{figure}

\begin{figure}[!ht]
\centering{
\includegraphics[bb=0 0 283 283,width=7.0cm]{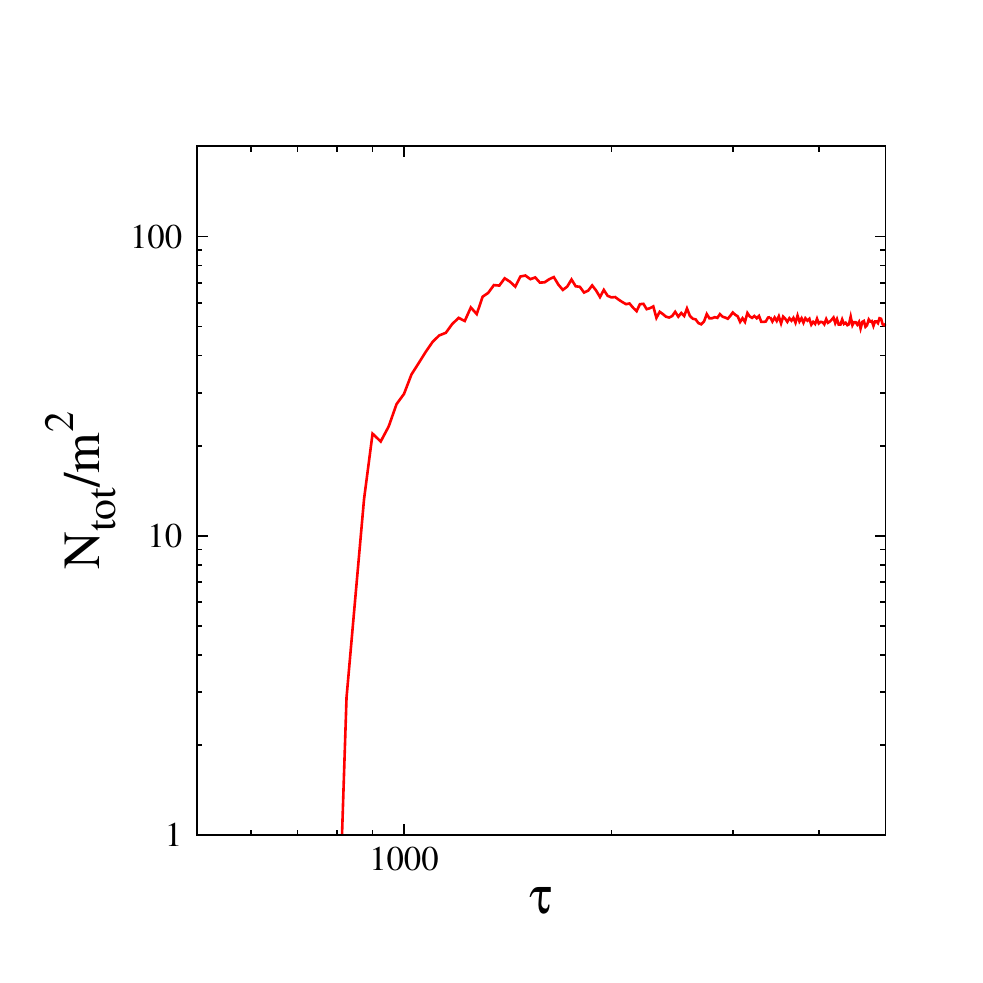}
}
\caption{The time evolution of the total number of Q-balls
 in the unit comoving area, for the parameter set 2D1 and
 $q_{\rm c} = 3\times 10^{-5}\unitq$.
} 
 \label{fig:ncount2D}
\end{figure}

Let us next consider the 2D system. The parameter sets used in our analysis are
shown in Table \ref{tab:2D}.  
We first show the result with the parameter set 2D1. 
The snapshots of the spatial charge distribution during the Q-ball formation are
shown in Fig.~\ref{fig:structure2D}. In this figure, the panels (a)--(d) correspond
to $\tau=725, 825, 875$ and $2000$, respectively, and the red indicates the
large positive charge and the black $q\sim 0$. 
We can clearly see the filamentary structure  just
before the Q-ball formation [see (a)(b)], which was noted in Ref.~\cite{Enqvist:2000cq}. Our point here is that the filamentary structure 
plays an important role in the Q-ball formation.
In order to help understanding the formation process, we classify the
Q-balls into those formed at intersections,
sides and voids of the filamentary structure,
in the order of the charges and the formation time.~\footnote{Note that this classification is obscure especially for small Q-balls; significant Q-ball interactions
are considered to occur locally, which makes it difficult to chase the origin of those Q-balls. However it is still useful to keep in mind such classification
because those small Q-balls do not significantly contribute to the peak of the charge distribution. }
At $\tau \sim 800$, large proto-Q-balls are formed at the intersections of
the filaments,  as one can see in (b). These proto-Q-balls will chiefly become the largest Q-balls forming the peak in the final distribution.
Shortly after the formation of the proto-Q-balls at the intersections,
the filaments start to be torn to smaller pieces, which will in the end
become Q-balls with smaller charges. In addition, small Q-balls are formed in the void regions of the
filamentary structure almost at the same time [see (c)].  The Q-balls formed from the sides and the voids will account for the tail of the charge distribution 
left to the peak. These classification may become vague after mergers and disruptions through collisions, but it is clear from Fig.~\ref{fig:structure2D}
that the filamentary structure strongly affects the spatial charge distribution even at late times. This argument is reinforced by the observation that major collisions are less 
frequent in the expanding background, as we have seen in the case of 1D (see Fig.~\ref{fig:trajectory}). Also we can explicitly see that 
the number of Q-balls is preserved soon after the Q-balls from the sides
and voids are formed, as shown in Fig.~\ref{fig:ncount2D}.

\begin{table}[!ht]
\begin{tabular}{c|ccccccc}
\hline
ID & grid size & $b$ & $dt$ & $K$ & $\epsilon$ \\
\hline\hline
2D1 & 512$^2$ & 2.5 & 0.005 & -0.1  & 1    \\
2D2 & 512$^2$ & 2.5 & 0.005 & -0.07 & 1    \\
2D3 & 512$^2$ & 3.0 & 0.005 & -0.04 & 1    \\
2D4 & 512$^2$ & 2.5 & 0.005 & -0.1  & 0.1  \\
2D5 & 512$^2$ & 2.5 & 0.005 & -0.1  & 0.01 \\
\hline
\end{tabular}
\caption{Numerical parameters for 2D simulations.}
\label{tab:2D}
\end{table}

\subsubsection{Charge distribution}
\label{subsubsec:charge2D}

\begin{figure}[!ht]
\centering{
\includegraphics[bb=0 0 283 283,width=7.0cm]{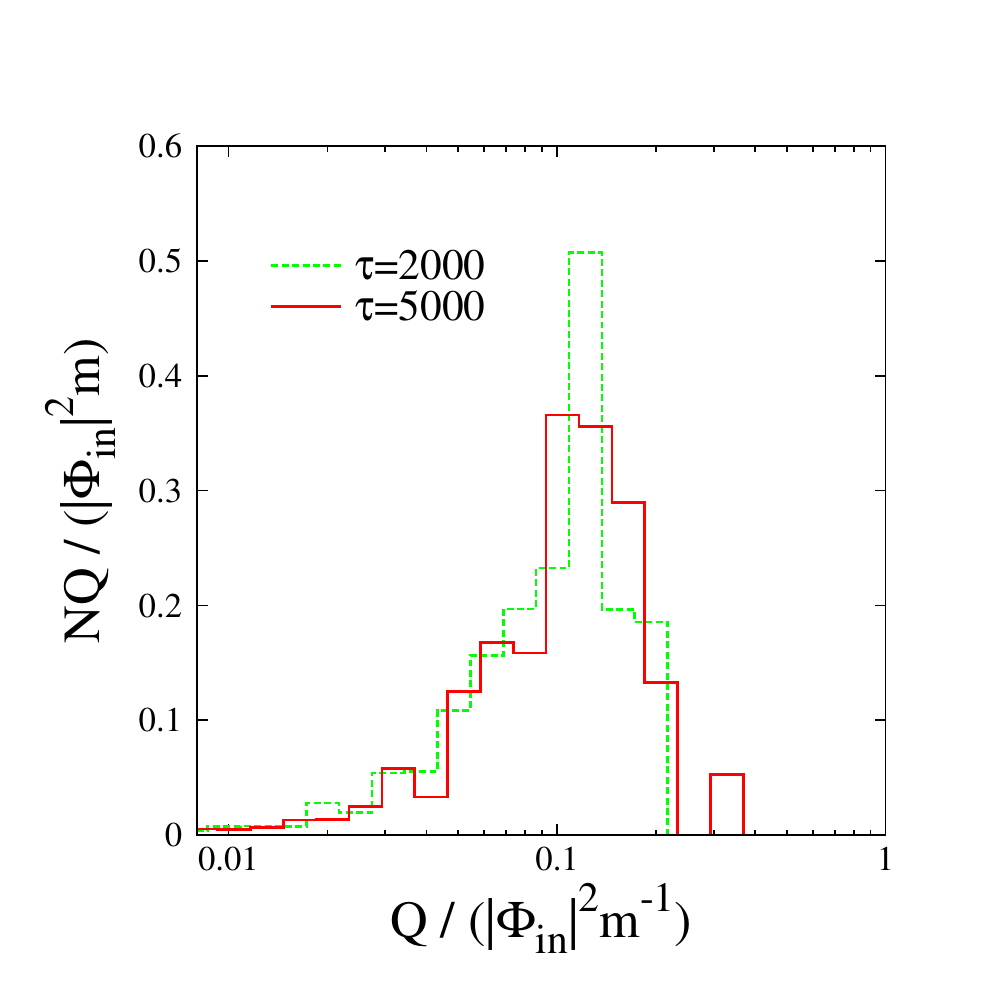}
}
\caption{Charge distributions of Q-balls at $\tau=2000$ (green dashed)
 and $\tau=5000$ (red solid), for
 the parameter set 2D1 and $q_{\rm c}=3\times 10^{-5}\unitq$. The
 vertical axis is the number of Q-balls with charge $Q$
 in a comoving area, $N(Q)$, multiplied by the charge. 
} 
 \label{fig:charge2D}
\end{figure}

Fig.~\ref{fig:charge2D} shows the charge distribution of Q-balls for the
2D1 simulation. We set the criterion $q_{\rm c}=3\times 10^{-5}\unitq$
for the identification of Q-balls.  The distribution does not
significantly change once the Q-balls of the three types are
formed. This is consistent with our observation that major collisions
are rare in the expanding universe. The distribution has a sharp peak at
$Q^{\rm 2D}_{\rm peak}\sim 0.1\unitQQ$ and the typical charge ranges over one
order of magnitude $0.03 \lesssim Q/(\unitQQ) \lesssim 0.2$. 

Let us compare the existing result with ours. The maximal charge
observed in the simulation of Ref.~\cite{Kasuya:2001hg} reads
$Q^{(KK)}_{\rm max}\sim 0.1 \unitQQ$ in our setup\footnote{Please notice
that in Ref.~\cite{Kasuya:2001hg}, the treatment of 1D and 2D Q-balls is
different from ours. The 2D Q-balls in Ref.~\cite{Kasuya2001hg} has an
infinitely long sylindrical shape in 3D space.}. This agrees very well
with the above peak value. 

We confirmed that the Q-balls formed at the intersections
account for the largest Q-balls, especially
those around the peak, while the Q-balls created from the sides and the voids form a long tail
toward smaller charges. {\it Thus we conclude that it is the filamentary
structure that determines the charge distribution, for an orbit with a
small ellipticity.} 

\subsubsection{Relationships among charge, energy and size}
\label{subsubsec:qes2D}

\begin{figure}[!ht]
\centering{
\includegraphics[bb=0 0 481 283,width=13cm]{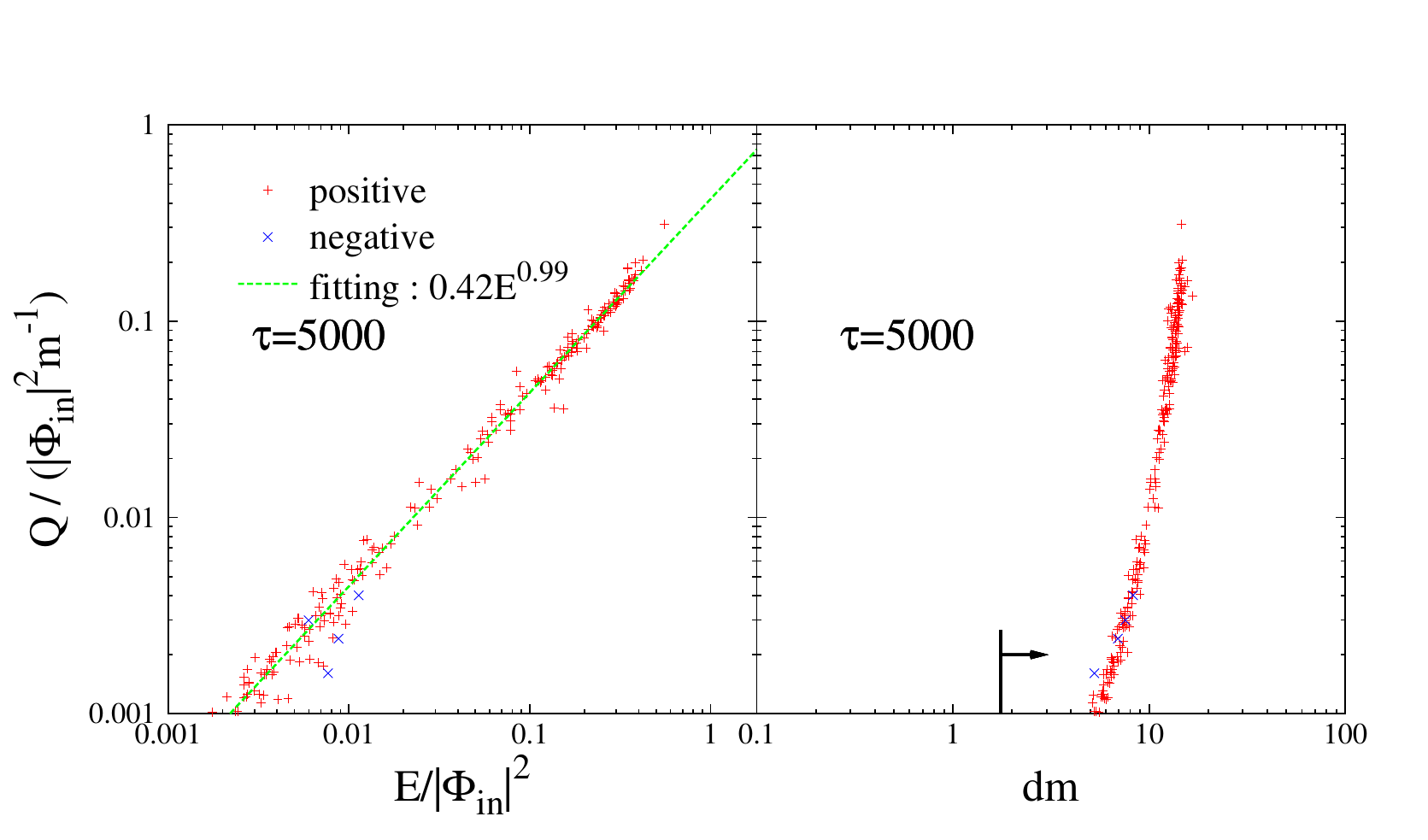}
}
\caption{The charge and the size (diameter $d=2R$) of Q-balls with
 parameter set 2D1 and $q_{\rm c}=3\times 10^{-5}\unitq$.
 The red `$+$' and the blue `$\times$' represent positive Q-balls and
 negative ones at $\tau=5000$, respectively.
 The green dashed line represents the fitting formula,
 $|Q|/(\unitQQ) = 0.42(E/\unitEE)^{0.99}$.
 The arrow standing on the horizontal axis represents the grid size at
 $\tau=5000$ ($\Delta x = 1.87 m^{-1}$).
} 
 \label{fig:qes2D}
\end{figure}

The gravity-mediation type Q-ball satisfies a relation between the
charge and the energy, $E\sim mQ$. To check this relation, we plot in
the left panel of Fig.~\ref{fig:qes2D} the charge and the energy of
the Q-balls in the 2D1 simulation. This figure shows that these
quantities are strongly correlated, having an almost linear relation,
$|Q|/(\unitQQ)\simeq 0.42(E/\unitEE)^{0.99}$ for $|Q|>10^{-3}\unitQQ$, 
as expected. Thus, it
confirms that our numerical algorithm for the formation and the
identification of Q-balls work well.  We note however that the
proportional coefficient is smaller than 1, which means that the orbit
of the AD field in the phase space is not completely circular. Strictly
speaking, those objects we have identified as Q-balls are not the Q-ball
solution but slightly excited states of the Q-balls~\cite{Enqvist:1999mv}. 
This excitation may be due to the Hubble friction at the early time
for the simulation. At the initial time, we set the initial field so
that $mQ/E \approx 1$ [see Eq.~(\ref{eq:init})]. The field, however,
cannot generally keep the circularity of its orbit in the phase space
while the Hubble friction works efficiently. Also, we will see the
dimensional effect on the proportional coefficient in
Sec.~\ref{subsubsec:qes3D}.

We have seen no further relaxation process toward the true Q-ball
solution in our numerical simulations. This is a limitation of our
analysis. However, since the coefficient is relatively close to $1$, we
expect that those objects will gradually approach the Q-ball solution in
a much longer time scale~\footnote{The relaxation time may be extremely
long due to an approximate adiabatic invariant~\cite{Kasuya:2002zs}.}, 
without significantly affecting the charge distribution obtained here. 

Before showing the relation between the charges and the sizes, we
mention the definition of {\it diameter} of a Q-ball in this paper.
The shape of a Q-ball identified by our algorithm is generically a
distorted disk (ball in the 3D system). So, our procedure to calculate
the {\it diameter} of a Q-ball is as follows. First, we measure the
area (the volume in the 3D system) of the region on the 
grid in which the charge density is larger than the criterion.
At this stage, the location of the boundary of a Q-ball is specified by
the linear interpolation using the two values of $q(t,\xx)$ at the
adjacent grid points between which $q(t,\xx)=q_{\rm c}$ is realized.
Then the definition of the diameter of a Q-ball in this paper
is the diameter of the disk (a ball) with the same area (volume) as
measured above, which is given in Eq.~(\ref{eq:diameter}). For details,
see Appendix B.

In the right panel of Fig.~\ref{fig:qes2D}, we plot the relation 
between the size (diameter $d=2R$) and the charge. 
The Q-ball solution
has a certain size determined from the instability band shown in
Eq.~(\ref{eq:inst}). We have found that all the Q-balls at least above
$Q>10^{-3}\unitQQ$ have an almost same size, $d\sim 10m^{-1}$. This is
another evidence that our numerical 
scheme correctly follows the Q-ball formation and collects the Q-balls.
The spatial resolution in this simulation is presented as an
arrow standing on the horizontal axis of Fig.~\ref{fig:qes2D}, being
much smaller than the typical size of a Q-ball. Hence the Q-balls are
successfully identified on the lattices with sufficient spatial resolution.

\subsubsection{K-dependence}
\label{subsubsec:K2D}

The strength of the instability is controlled by the coefficient of the
one-loop corrections, $K$. For a larger value of $|K|$, the instability gets stronger and
the Q-ball size $R\sim |K|^{-1/2}m^{-1}$ becomes smaller. 
Thus, the final Q-ball charge distribution may depend on the value of $K$.

\begin{figure}[!ht]
\centering{
\includegraphics[bb=0 0 263 263, width=8cm]{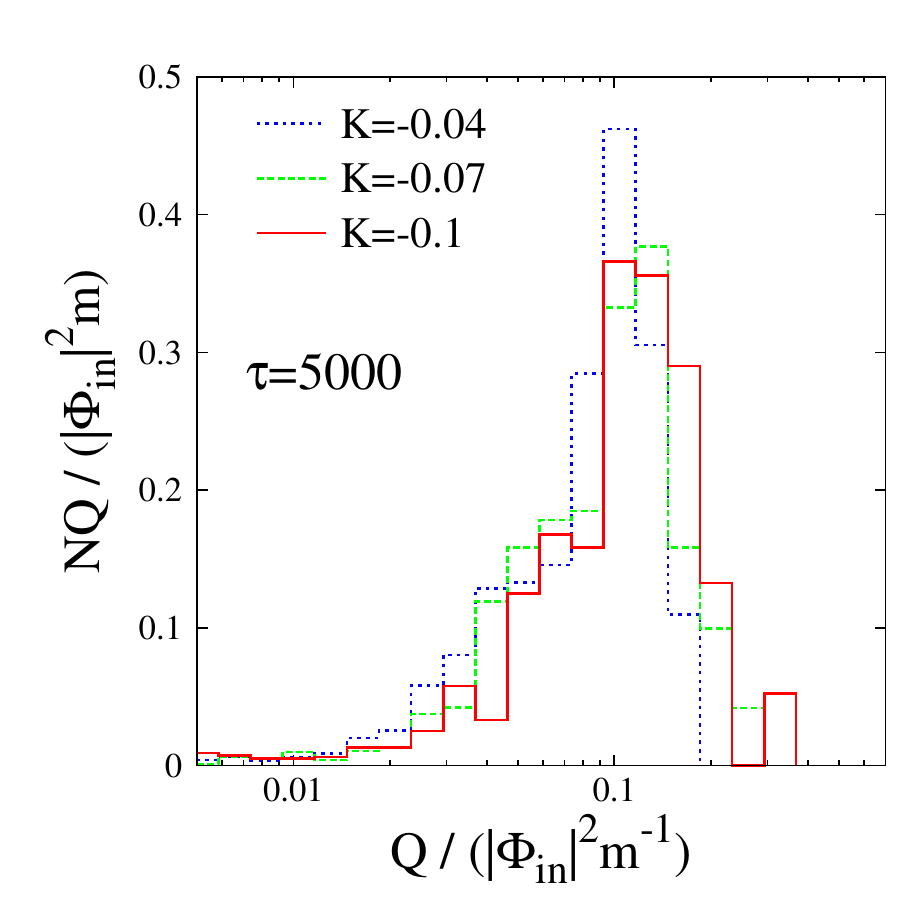}
}
\caption{Charge distributions of Q-balls with
$K=-0.1, q_{\rm c}=3\times 10^{-5}\unitq$ [2D1] (red solid),
$K=-0.07, q_{\rm c}=2\times 10^{-5}\unitq$ [2D2] (green dashed), and
$K=-0.04, q_{\rm c}=1.2\times 10^{-5}\unitq$ [2D3] (blue dotted).
} 
 \label{fig:Kcharge2D}
\end{figure}

Figure \ref{fig:Kcharge2D} shows the charge distributions at $\tau = 5000$ with 
$K=-0.1, -0.07$ and $-0.04$, corresponding to 2D1, 2D2 and 2D3,
respectively, in Table \ref{tab:2D}. Although the charge distribution seem to 
shift toward larger values of Q as $|K|$,  no significant difference was
observed. 

\begin{figure}[!ht]
\centering{
\includegraphics[bb=0 0 283 283,width=8cm]{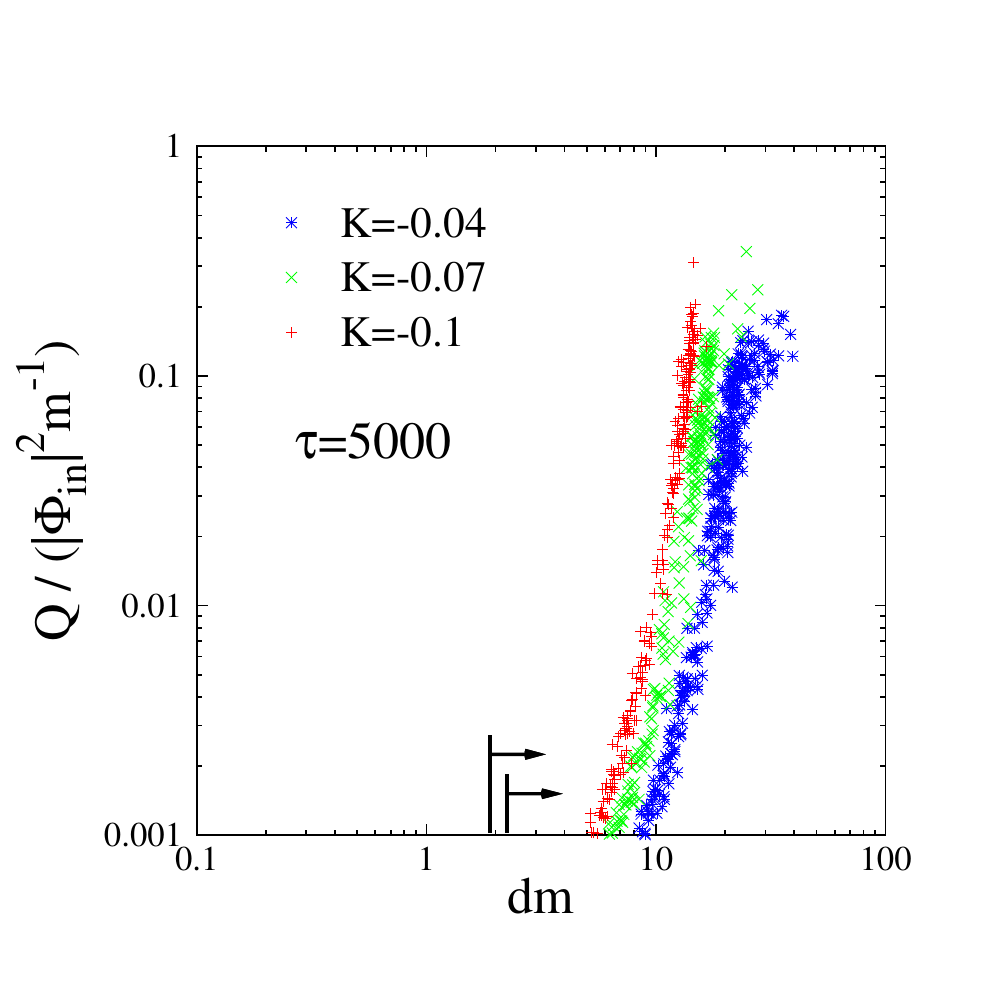}
}
\caption{The charge and the size (diameter $d=2R$) of Q-balls with
$K=-0.1, q_{\rm c}=3\times 10^{-5}\unitq$ [2D1] (red `$+$'),
$K=-0.07, q_{\rm c}=2\times 10^{-5}\unitq$ [2D2] (green `$\times$'), and
$K=-0.04, q_{\rm c}=1.2\times 10^{-5}\unitq$ [2D3] (blue `$*$'),
The arrows standing on the horizontal axis represent the grid size at
 $\tau=5000$ for $K=-0.1, 0.07$ simulations (tall one,
  $\Delta x = 1.87 m^{-1}$) and for $K=-0.04$ simulation (short one,
 $\Delta x = 2.25 m^{-1}$).
} 
 \label{fig:Kqs2D}
\end{figure}

The size (diameter $d=2R$) and the charge of Q-balls in simulations
2D1-2D3 are shown in Fig.~\ref{fig:Kqs2D}.
Focusing on the large Q-balls with $Q=Q^{\rm 2D}_{\rm peak} \pm 10\%$ in each
result~\footnote{The reason why we select the large ones is that the
error in measuring the size is considered to be minimized. Smaller
Q-balls tend to have smaller radius in our Q-ball identification
algorithm due to the presence of the ambient background fluctuations.},
we can read the typical sizes as 
$d=2R\sim 14m^{-1}, 17m^{-1}, 24m^{-1}$ for $K=-0.1, -0.07, -0.04$, 
respectively. Thus we confirmed that the size approximately scales as
$R\sim |K|^{-1/2}m^{-1}$. More specifically, we obtain 
$R \sim 2.3|K|^{-1/2}m^{-1}$ at $\tau=5000$. 

\subsubsection{Initial elliptic orbit}
\label{subsubsec:elliptic2D}

So far we have concentrated on the cases with $\epsilon = 1$ [see
Eq.~(\ref{eq:init}) for its definition]. If the parameter $\epsilon$ is
smaller than $1$, the Q-ball formation and the charge distribution are
significantly changed. However, the filamentary structure is still
essential to understand the final distribution. 

For $\epsilon \ll 1$,  the initial orbit of the AD field becomes
elliptic. It has been known that negative Q-balls, namely, Q-balls with
a negative charge, appear in this
case~\cite{Kasuya:2000wx,Enqvist:2000cq}. This is because the AD field
cannot grow into the Q-balls while keeping the energy-to-charge ratio
$E/mQ \sim 1/\epsilon \gg1$. (Recall that the Q-ball solution satisfies
$E\sim mQ$.) Thus, the AD field must somehow discard its excessive
energy to reach the Q-ball solution, while the total charge is
conserved. This can only be achieved by changing the ratio locally; in
reality, this results in producing negative Q-balls. 

\begin{figure}[!ht]
\begin{tabular}{cc}
\begin{minipage}{6cm}
\includegraphics[bb=0 0 503 479,width=6cm]{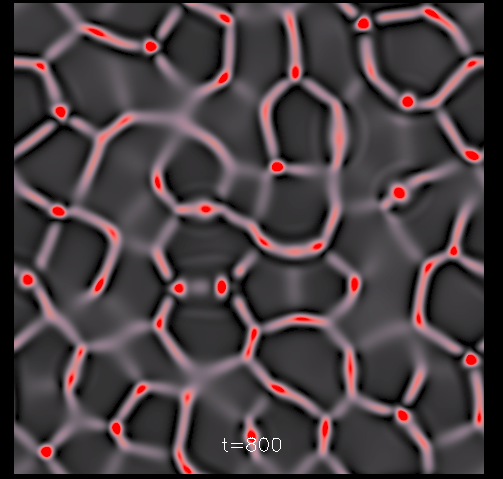}
(a) $\tau=800$
\end{minipage}
\begin{minipage}{6cm}
\includegraphics[bb=0 0 503 479,width=6cm]{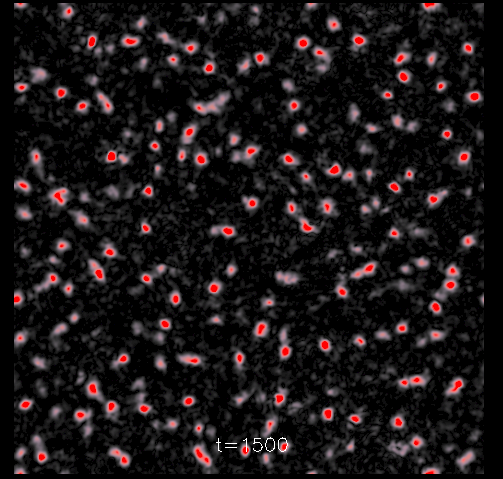}
(b) $\tau=1500$
\end{minipage}
\\
\begin{minipage}{6cm}
\includegraphics[bb=0 0 503 479,width=6cm]{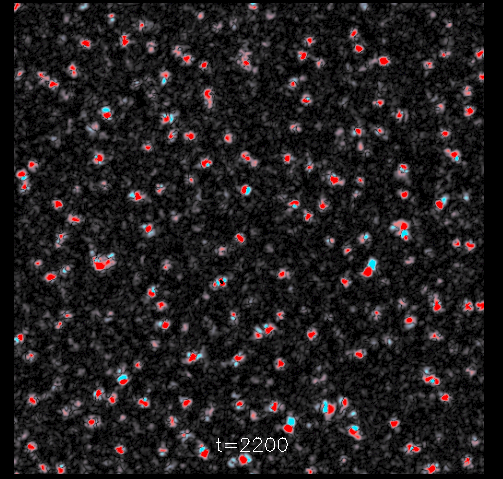}
(c) $\tau=2200$
\end{minipage}
\begin{minipage}{6cm}
\includegraphics[bb=0 0 503 479,width=6cm]{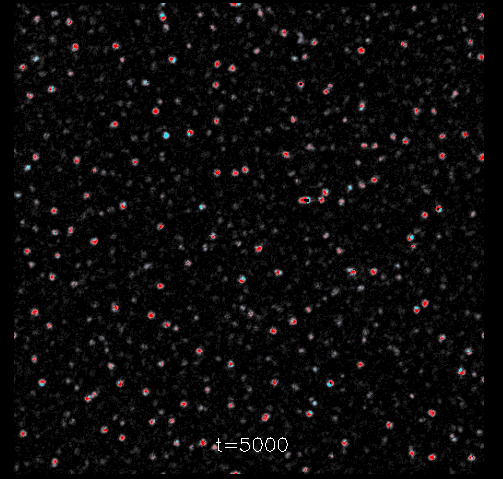}
(d) $\tau=5000$
\end{minipage}
\end{tabular}
\caption{The color map of the charge density for $\epsilon=0.1$. The
 red represents the positive charge, the cyan the negative one and the
 black $q\sim 0$.
} 
 \label{fig:eps01_2D}
\end{figure}

\begin{figure}[!ht]
\begin{tabular}{cc}
\begin{minipage}{6cm}
\includegraphics[bb=0 0 503 479,width=6cm]{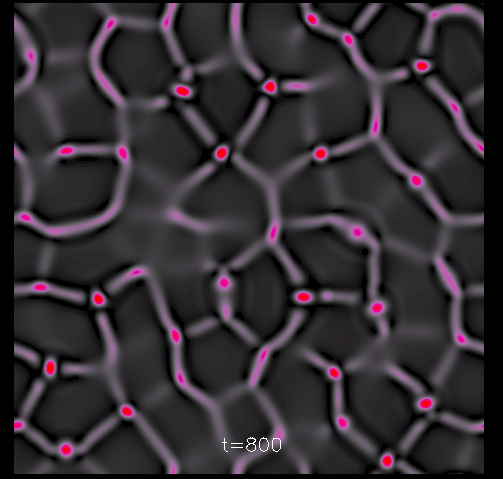}
(a) $\tau=800$
\end{minipage}
\begin{minipage}{6cm}
\includegraphics[bb=0 0 503 479,width=6cm]{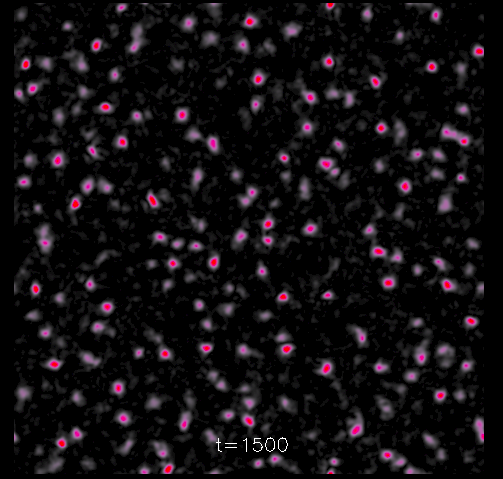}
(b) $\tau=1500$
\end{minipage}
\\
\begin{minipage}{6cm}
\includegraphics[bb=0 0 503 479,width=6cm]{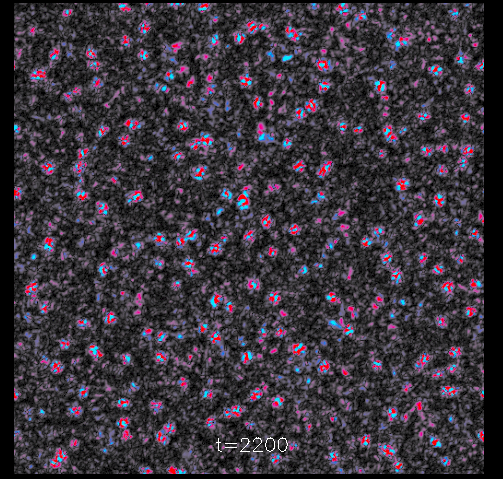}
(c) $\tau=2200$
\end{minipage}
\begin{minipage}{6cm}
\includegraphics[bb=0 0 503 479,width=6cm]{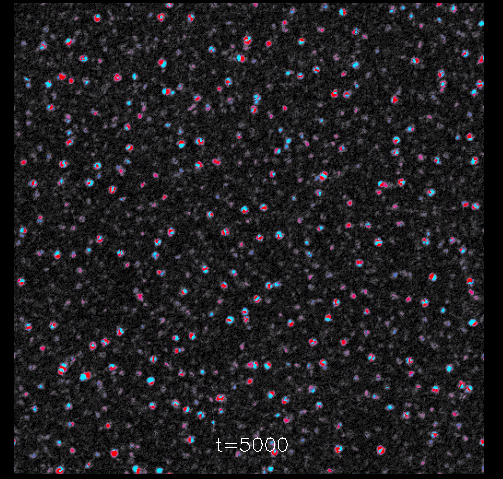}
(d) $\tau=5000$
\end{minipage}
\end{tabular}
\caption{The color map of the charge density for $\epsilon=0.01$. The
 red represents the positive charge, the cyan the negative one and the
 black $q\sim 0$. 
} 
 \label{fig:eps001_2D}
\end{figure}

We have performed numerical simulations with $\epsilon = 0.1$
(2D4) and $\epsilon=0.01$ (2D5). The snapshots of the spatial charge
distribution at $\tau = 800, 1500, 2200$ and $5000$ are shown in
Figs.~\ref{fig:eps01_2D} and \ref{fig:eps001_2D}, respectively. 
In these figures, the red represents the positive Q-balls and the cyan
the negative ones. Let us first explain how the formation proceeds in
the case of $\epsilon = 0.1$. As in the case of $\epsilon = 1$, the
filamentary structure appears first [see the panel (a)]. As the field
grows, the filaments are torn to smaller pieces and the proto-Q-balls
are formed from intersections,
sides and voids of the filaments. Similarly to
the case of $\epsilon = 1$, the largest Q-balls are those from the
intersections.
At this stage, we call these Q-balls {\it first-generation Q-balls}.
Almost all the first-generation Q-balls have a positive charge, as can
be seen in the panel (b). Note that the first-generation Q-balls are in
a highly excited state in a sense that they have large $E/Q$
ratio. Around $\tau\sim 2200$ [the panel (c)], negative Q-balls (cyan)
begin to be formed. The point to be emphasized here is that the negative
(as well as positive) Q-balls, especially larger ones,  appear around
the existing first-generation Q-balls. 
If one takes a closer look, one may notice that a pair of positive and
negative Q-balls is produced from one first-generation Q-ball. This
observation clearly shows that the highly excited first-generation
Q-balls release the excessive energy in a form of positive and negative
Q-ball pairs. At $\tau = 5000$ [the panel (d)], the relaxation process
has been almost finished. We call these Q-balls {\it second-generation
Q-balls} to distinguish them from the first-generation
Q-balls. We can 
see some isolated negative second-generation Q-balls, but most of them
are still paired with the positive ones.  

In the case of $\epsilon = 0.01$, the evolution is quite similar until
the first-generation Q-balls are formed (the panel (b) in
Fig.~\ref{fig:eps001_2D}). The production of the negative Q-balls are
more violent as can be seen from the panel (c). We can see multiple
positive and negative Q-balls are produced from one first-generation
Q-ball. This should be contrasted to the previous case that a pair of
positive and negative Q-balls is produced. Most of the large negative and
positive Q-balls remain paired even at $\tau = 5000$. Note that the
largest (positive or negative) Q-balls are produced from the largest
first-generation Q-balls, which chiefly arise at the intersections
of the
filamentary structure. In the void regions, many small Q-balls appear in
comparison with the case of $\epsilon = 0.1$. This may be induced by the
energy released from the first-generation Q-balls in the violent
relaxation process. 

\begin{figure}[!ht]
\centering{
  \includegraphics[bb=0 0 510 481, width=13cm]{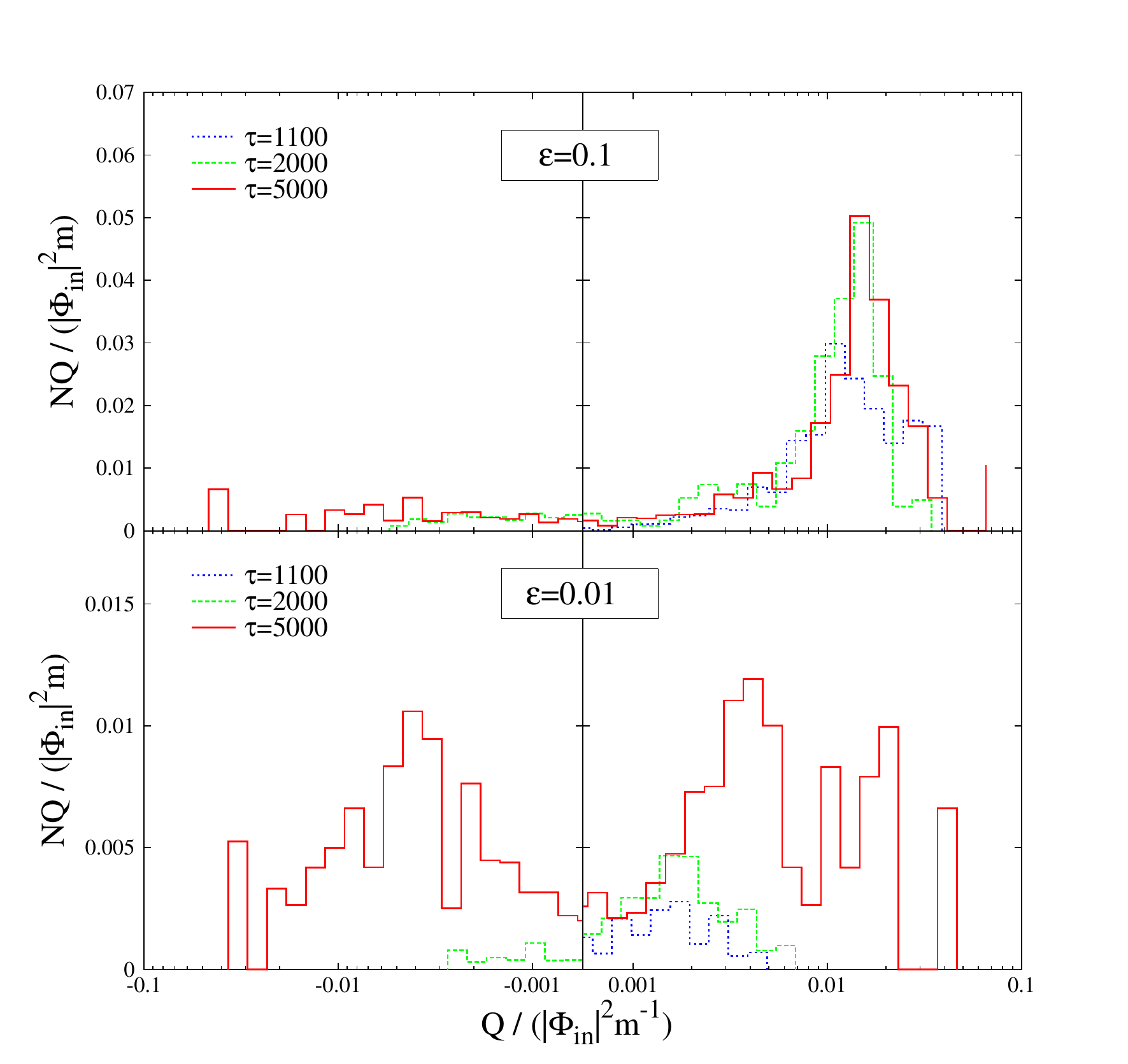}
}
\caption{Charge distributions of Q-balls with 
 $q_{\rm c}=\pm3\times 10^{-5}\epsilon\unitq$. The upper panel is the
 case with $\epsilon=0.1$ [2D4], and the lower one is $\epsilon=0.01$ [2D5].
 The left panels are distributions of negative Q-balls and the right
 panels are those of positive Q-balls. 
} 
 \label{fig:eps_charge2D}
\end{figure}

We expect that the filamentary structure also affects  the final charge
distribution, since it played an essential role in the Q-ball formation.
Fig.~\ref{fig:eps_charge2D} shows the charge distributions of positive and
negative Q-balls. The upper panels are the result with $\epsilon=0.1$, and
the lower ones that with $\epsilon=0.01$. We take the critical charge
$q_{\rm c}$ being proportional to $\epsilon$ as 
$q_{\rm c}=\pm3\times 10^{-5}\epsilon\unitq$, since the initial charge
density is proportional to $\epsilon$. In the case of $\epsilon=0.1$,
most Q-balls are positive, and negative Q-balls are subdominant. The peak
charge is approximately $Q^{\rm 2D}_{\rm peak}\sim 0.015\unitQQ$ which indicates
the scaling as $Q^{\rm 2D}_{\rm peak}\propto\epsilon$. (Recall that the peak
charge is $Q^{\rm 2D}_{\rm peak}\sim 0.1\unitQQ$ for $\epsilon = 1$. See
Fig.~\ref{fig:charge2D}.) On the other hand, in the case of
$\epsilon=0.01$ where the initial orbit is highly eccentric, almost the same
number of positive and negative Q-balls are produced after 
$\tau \sim 2000$.  The peak charge is approximately $2\times 10^{-3}$
until $\tau \lesssim 2000$ which agrees with the scaling. However, the
scaling of the peak charge is no longer valid at $\tau =
5000$~\cite{Kasuya:2001hg}.

\begin{figure}[!ht]
\centering{
\includegraphics[bb=0 0 471 283,width=13cm]{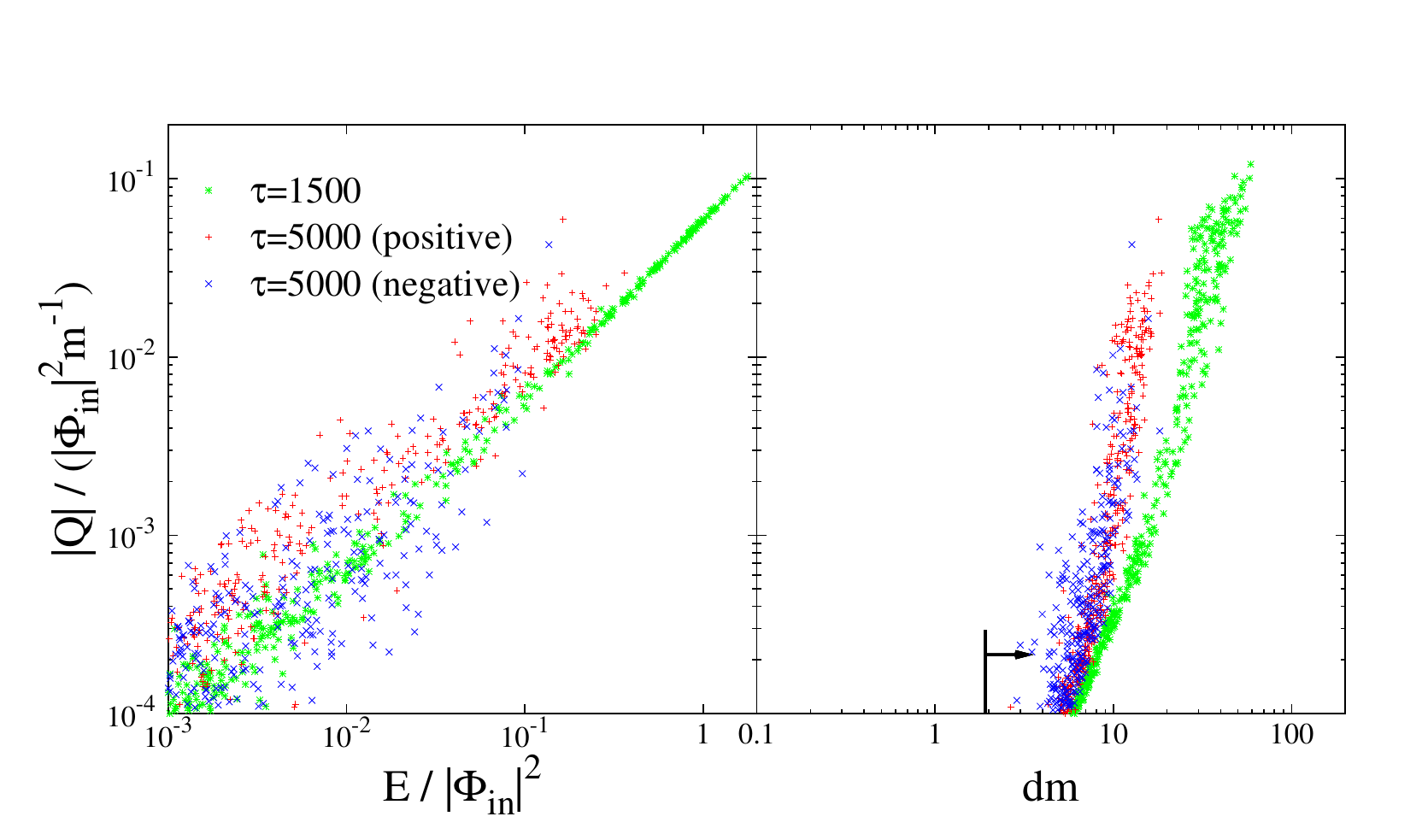}
}
\caption{The charge and the size (diameter $d=2R$) of Q-balls with
 $\epsilon=0.1, q_{\rm c}=3\times 10^{-6}\unitq$ [2D4]. 
 The green `*' represents those of positive Q-balls at $\tau=1500$.
 The red `$+$' and the blue `$\times$' represent positive Q-balls and
 negative ones at $\tau=5000$, respectively.
 The arrows standing on the horizontal axis represent the grid size at
 $\tau=5000$ ($\Delta x = 1.87 m^{-1}$).
} 
 \label{fig:eps01_qes2D}
\end{figure}

\begin{figure}[!ht]
\centering{
\includegraphics[bb=0 0 481 283,width=13cm]{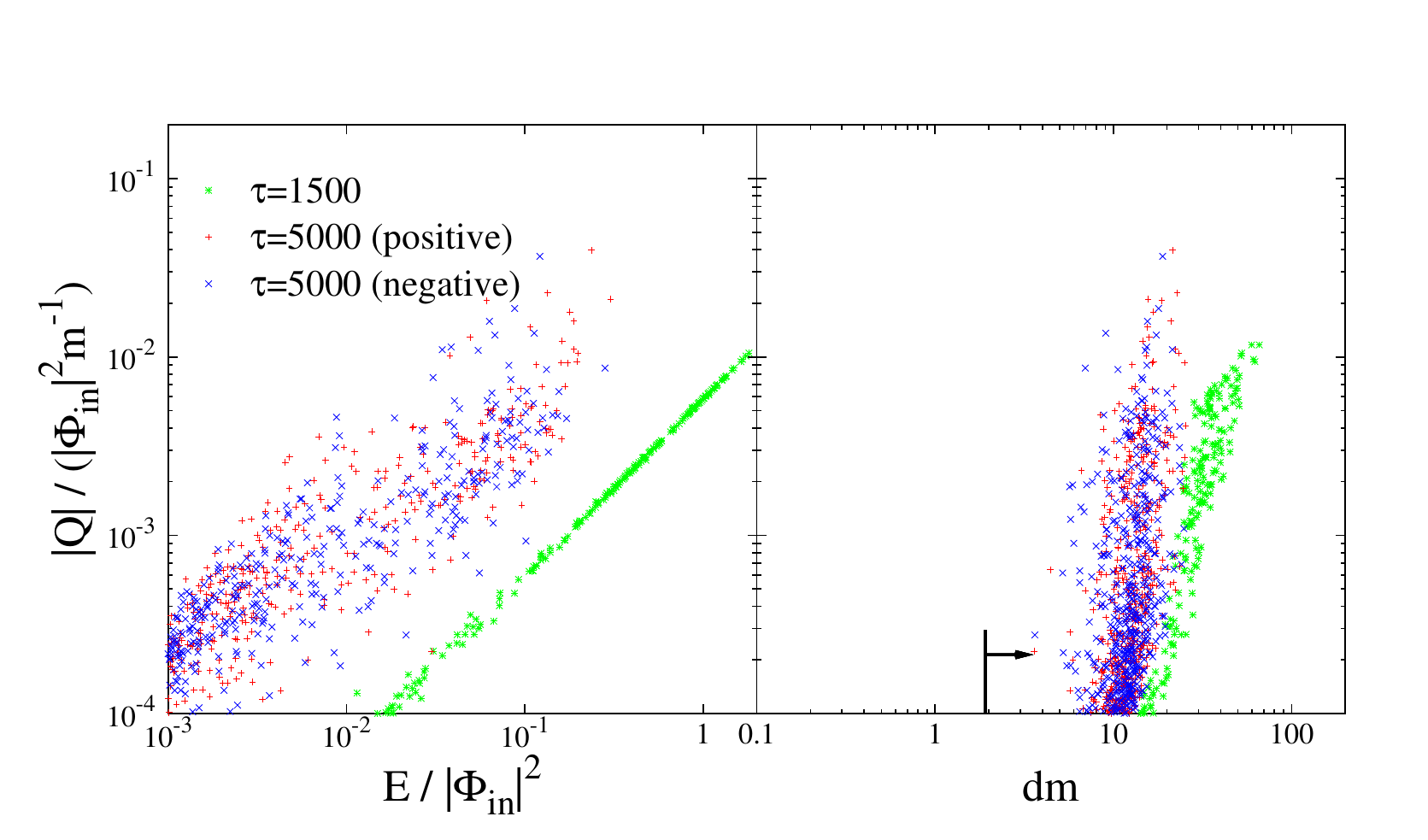}
}
\caption{The charge and the size (diameter $d=2R$) of Q-balls with
 $\epsilon=0.01, q_{\rm c}=3\times 10^{-7}\unitq$ [2D5]. 
 The green `*' represents those of positive Q-balls at $\tau=1500$.
 The red `$+$' and the blue `$\times$' represent positive Q-balls and
 negative ones at $\tau=5000$, respectively.
 The arrows standing on the horizontal axis represent the grid size at
 $\tau=5000$ ($\Delta x = 1.87 m^{-1}$).
} 
 \label{fig:eps001_qes2D}
\end{figure}

In order to study how the Q-balls in an excited state approach toward the
Q-ball solution, we show the energy and the charge of the Q-balls in the
left panels of Fig.~\ref{fig:eps01_qes2D} and \ref{fig:eps001_qes2D}, and
the relation between the size (diameter) and the charge in the right
panels. In the figures, the green symbols (`*') represent the energy and 
charge of the first-generation Q-balls at $\tau=1500$, implying
$E/mQ\sim 1/\epsilon$. Besides, the dependence of the size on the charge
is somewhat large. These facts indicate the first-generation Q-balls
shown here are highly excited states from the static Q-ball solution.
The red symbols (`$+$') and the blue ones (`$\times$') represent the
positive and the negative Q-balls at $\tau=5000$, respectively.
We can see that the energy-to-charge ratio decreases as time goes,
while roughly keeping the proportionality between $E$ and $Q$, and the sizes
become almost universal. The Q-balls seem to stop evolving after the
ratio $E/mQ$ reaches $O(10)$. These figures manifestly show the
relaxation process from the Q-balls in an excited state to a state with
a lower value of $E/Q$. 

The relaxation process of an excited state of Q-balls is a highly
non-linear phenomenon, which can be studied only by numerical
simulations. From our results shown above, we have reached conclusion
that Q-balls in an excited state, namely having a large $E/mQ \gg1$,
continue to transform the excessive energy into multiple (positive and
negative) Q-balls, until the $E/mQ$ ratio becomes
as small as $O(10)$\footnote{The Q-balls might have some amount of the
angular momentum. However, we looked into the kinetic, potential and
gradient energy of the Q-balls and found that the angular momentum is
not the main source for the large $E/mQ$ ratio.}. 
In the case of  $E/mQ \gg 10 $, or equivalently, $\epsilon \ll 0.1$,
therefore, both positive and negative Q-balls are produced in the
relaxation process, forming an almost symmetric charge distribution.

In the case with $\epsilon=1$, we pointed out that the ratio $E/mQ$ is
larger than 1 due to the Hubble friction. In the present case,
however, the Hubble friction is no longer relevant at the formation of
the second-generation Q-balls. Hence the fact that the
second-generation Q-balls have somewhat large $E/mQ$ may be due to the
highly non-linear interaction during the secondary formation process.
To find out the reason, more detailed investigations on the secondary
formation process are required, which is beyond the scope of this paper.


\subsection{3D}
\label{subsec:3D}

\begin{table}[!ht]
\begin{tabular}{c|ccccccc}
\hline
ID & $\#$ of grid & $b$ & $dt$ & $K$ & $\epsilon$ \\
\hline\hline
3D1 & 128$^3$ & 1.0 & 0.005 & -0.1  & 1.0  \\
3D2 & 128$^3$ & 1.0 & 0.005 & -0.07 & 1.0  \\
3D3 & 128$^3$ & 1.6 & 0.005 & -0.04 & 1.0  \\
3D4 & 128$^3$ & 1.0 & 0.005 & -0.1  & 0.1  \\
3D5 & 128$^3$ & 1.0 & 0.005 & -0.1  & 0.01 \\
\hline
\end{tabular}
\caption{Numerical parameters for 3D simulations.}
\label{tab:3D}
\end{table}

\subsubsection{Filamentary structure}
\label{subsubsec:structure3D}

In this final section, we move on to 3D simulations. 
The parameter sets used are tabulated in Table \ref{tab:3D}.

\begin{figure}[!ht]
\begin{tabular}{cc}
\begin{minipage}{6cm}
\includegraphics[bb=0 0 640 640,width=6.0cm]{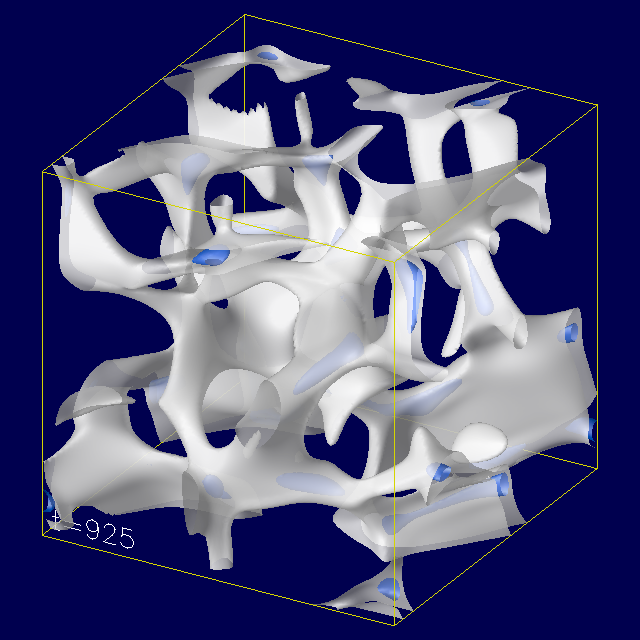}
(a) $\tau=925$
\end{minipage}
\begin{minipage}{6cm}
\includegraphics[bb=0 0 640 640,width=6.0cm]{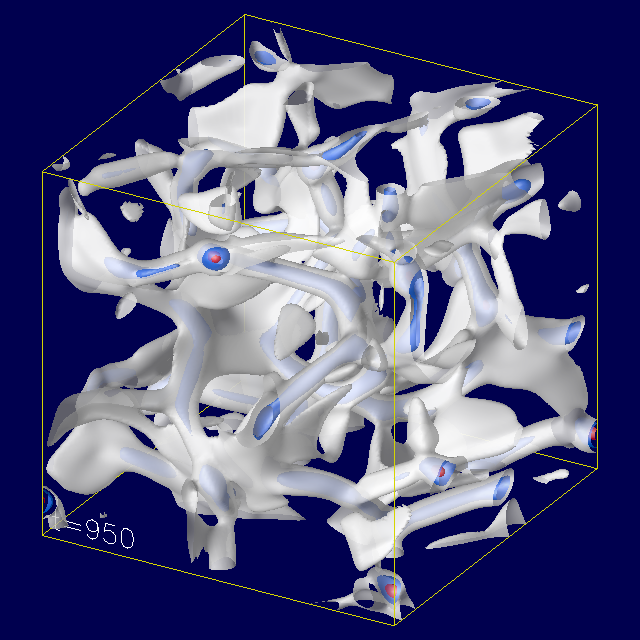}
(b) $\tau=950$
\end{minipage}
\\
\begin{minipage}{6cm}
\includegraphics[bb=0 0 640 640,width=6.0cm]{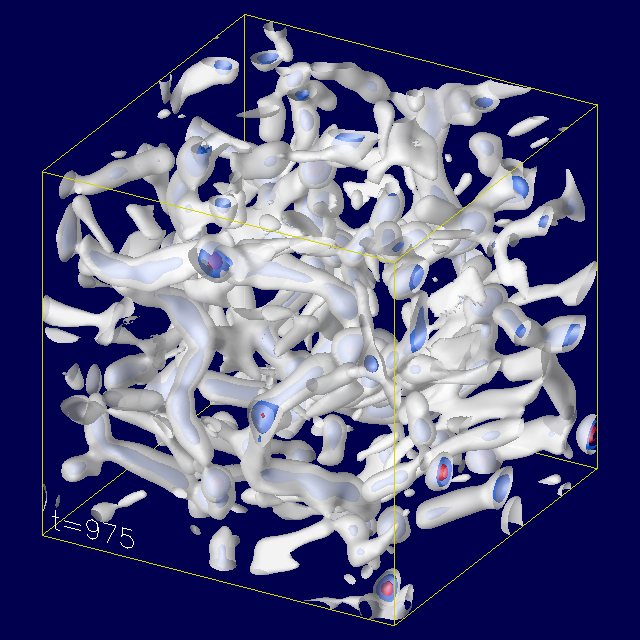}
(c) $\tau=975$
\end{minipage}
\begin{minipage}{6cm}
\includegraphics[bb=0 0 640 640,width=6.0cm]{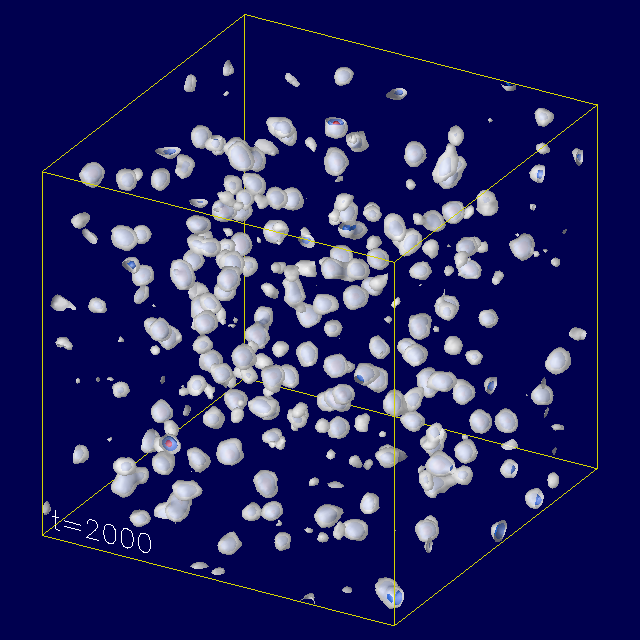}
(d) $\tau=2000$
\end{minipage}
\end{tabular}
\caption{The isosurfaces of charge density with 
 $q_{\rm c}=2\times 10^{-6}\unitq$ (white), 
 $q_{\rm c}=1\times 10^{-5}\unitq$ (blue), 
 and $q_{\rm c}=5\times 10^{-5}\unitq$ (red). 
For these visualizations, we used OpenDX.
} 
 \label{fig:structure3D}
\end{figure}

\begin{figure}[!ht]
\centering{
\includegraphics[bb=0 0 283 283, width=8.0cm]{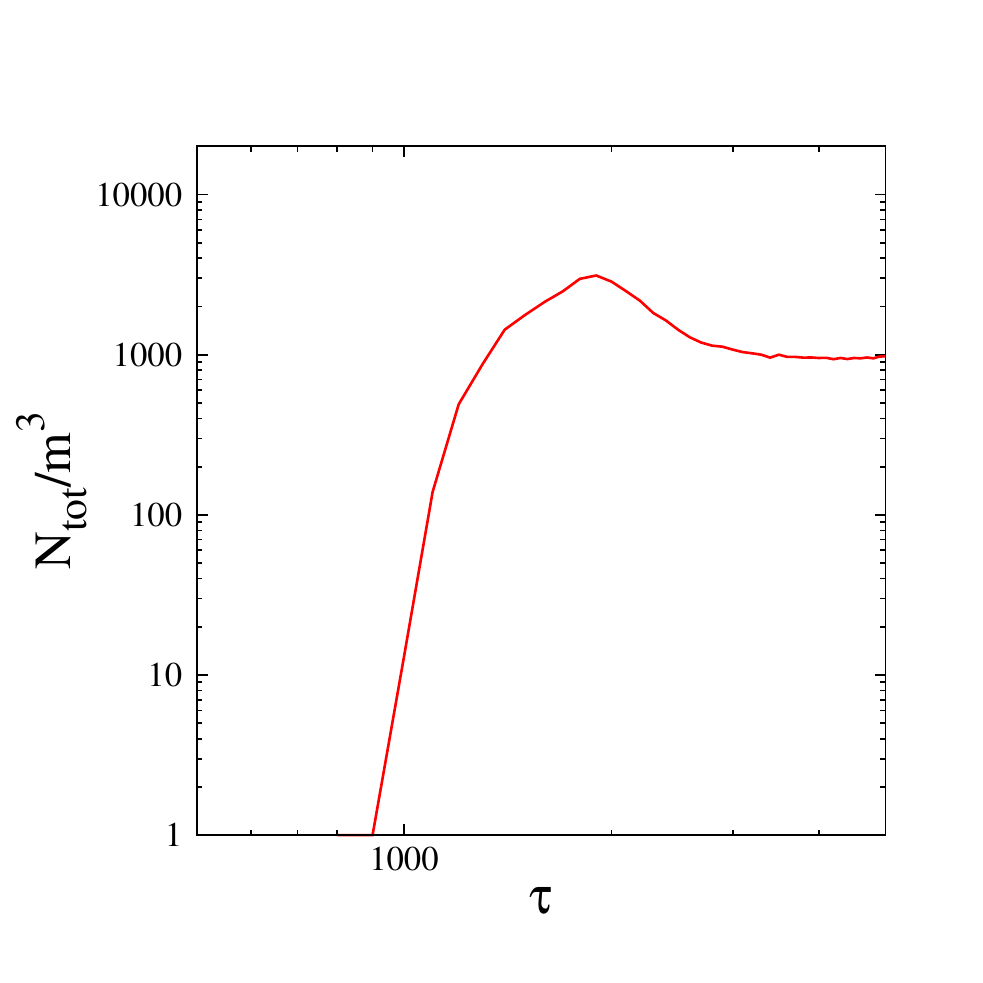}
}
\caption{The time evolution of the total number of
 Q-balls in the unit comoving volume, for the parameter set 3D1
 and $q_{\rm c}=1.8\times 10^{-7}\unitq$.
} 
 \label{fig:ncount3D}
\end{figure}

The snapshots of the spatial charge distribution 
with the parameter set 3D1 are shown in Fig.~\ref{fig:structure3D}. 
The surfaces in the figures represent the isodensity surfaces of the
charge density $q=2\times 10^{-6} \unitq$ (white), 
$q=1\times 10^{-5} \unitq$ (blue), and $q=5\times 10^{-5} \unitq$ (red). 
The panels (a)--(d) correspond to $\tau=925, 950, 975$ and $2000$,
respectively. As we have seen in
the 2D results, we observed the filamentary structure that appears
just before the Q-ball formation, and at each intersection,
proto-Q-balls are formed represented by blue surfaces in the panel (a). 
At $\tau=950$, Q-balls appear also on the sides, and at the same time,
the Q-balls at the intersections grow which are represented by red
surfaces. In the next panel (c), the filaments start to be torn to small
pieces which eventually become small Q-balls. This panel shows manifestly
that the Q-balls formed at the intersections hold large charges also
in the 3D system, and thus they mainly contribute to the charge
distribution. At the end of the simulation, $\tau=5000$, we observed
that the number of Q-balls in a unit volume becomes $N=10^3$
shown in Fig.~\ref{fig:ncount3D}.

The authors of Ref.~\cite{Kasuya:2000wx} concluded that only a few
number of Q-balls hold almost all charge of the whole simulation box.
In contrast to this, our result indicates that, although the box size is
smaller than that used in their simulations, there are quite large
number of Q-balls with almost same size. This discrepancy may come from
the difference of the spatial resolutions. In fact, while the resolution
of the smaller box in their simulation is given by $\Delta x\sim 0.05m^{-1}$
at the initial time, the one in our simulation is $\Delta x\sim 0.008m^{-1}$.
Moreover, even at the formation epoch, the resolution is given by
$\Delta x\sim a(1000)\times m^{-1}/128\sim 1.0m^{-1}$, and thus Q-balls are
resolvable at the epoch.
In compensation for the fine resolution, the box size in our simulation
is smaller than the past one. Nevertheless, 
Fig.~\ref{fig:structure3D} indicates that the Q-ball formation caused by
the collapse of the filamentary structure proceeds without being
affected by the finite volume effect. 


\subsubsection{Charge distribution}
\label{subsubsec:charge3D}

\begin{figure}[!ht]
\centering{
\includegraphics[bb=0 0 283 283,width=8.0cm]{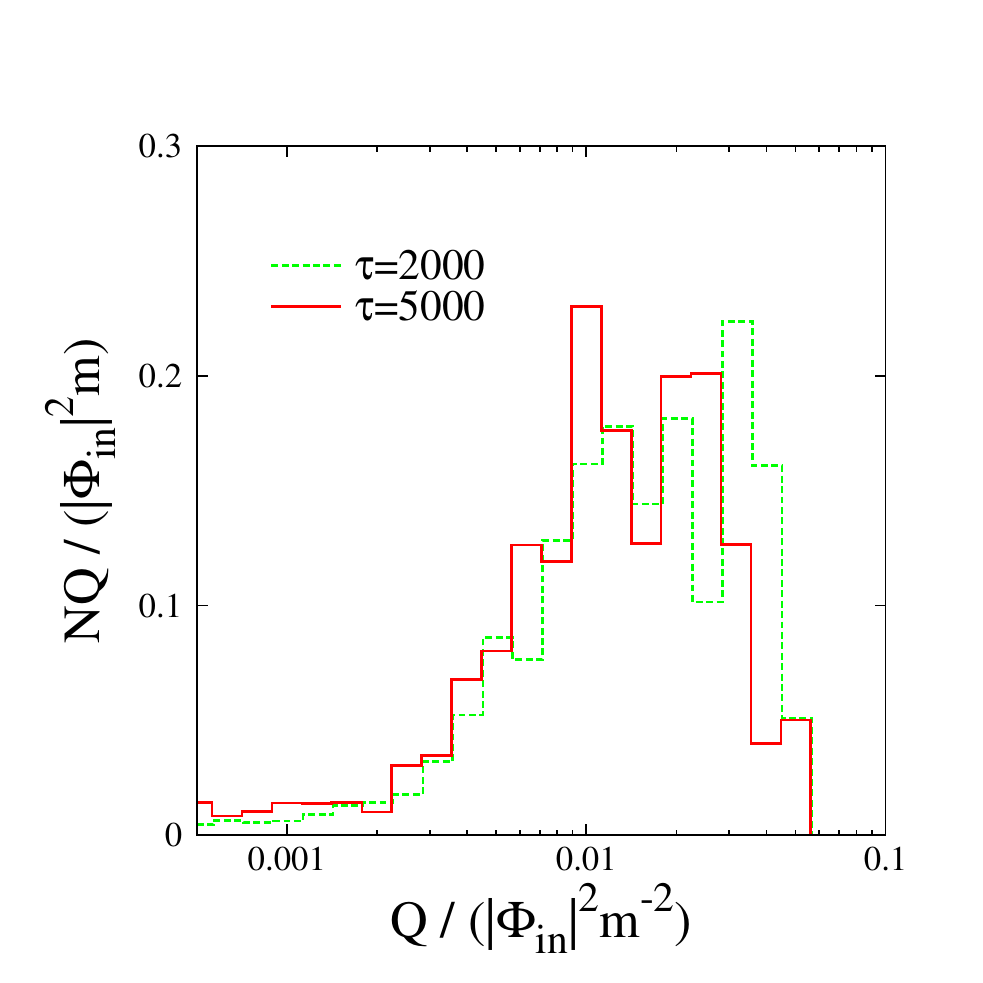}
}
\caption{Charge distributions of Q-balls at $\tau=2000$ (green dashed)
 and $\tau=5000$ (red solid), for
 the parameter set 3D1 and $q_{\rm c}=1.8\times 10^{-7}\unitq$. The
 vertical axis is the number of Q-balls with charge $Q$
 in a comoving volume, $N(Q)$, multiplied by the charge. 
} 
 \label{fig:charge3D}
\end{figure}

Fig.~\ref{fig:charge3D} shows the charge distribution of Q-balls for the 
3D1 simulation. We set the criterion, $q_c=1.8\times 10^{-7}\unitq$.
Basically, the Q-ball charges have a distribution similar to the
2D cases, and those mainly contributing to the total charge range over an
order of magnitude, namely, $2\times 10^{-3} \lesssim Q/(\unitQQQ)
\lesssim 5\times 10^{-2}$.
In Sec.~\ref{subsec:fitting}, we will perform the fitting of this distribution
in consideration of the Poisson error of the number of Q-balls in each
bin. As a result, 
the typical Q-ball charge is
$Q^{\rm 3D}_{\rm peak}\simeq 1.9\times 10^{-2} \unitQQQ$, 
which is somewhat larger than the previous
result obtained by Kasuya et al.~\cite{Kasuya:2000wx}
where they have concluded
$Q^{(KK)}_{\rm max}=1.2\times 10^{-2} \unitQQQ$ for $K=-0.01$, when
expressed in terms of our terminology.

\subsubsection{Relationships among charge, energy and size}
\label{subsubsec:qes3D}

\begin{figure}[!ht]
\centering{
\includegraphics[bb=0 0 481 283,width=16.0cm]{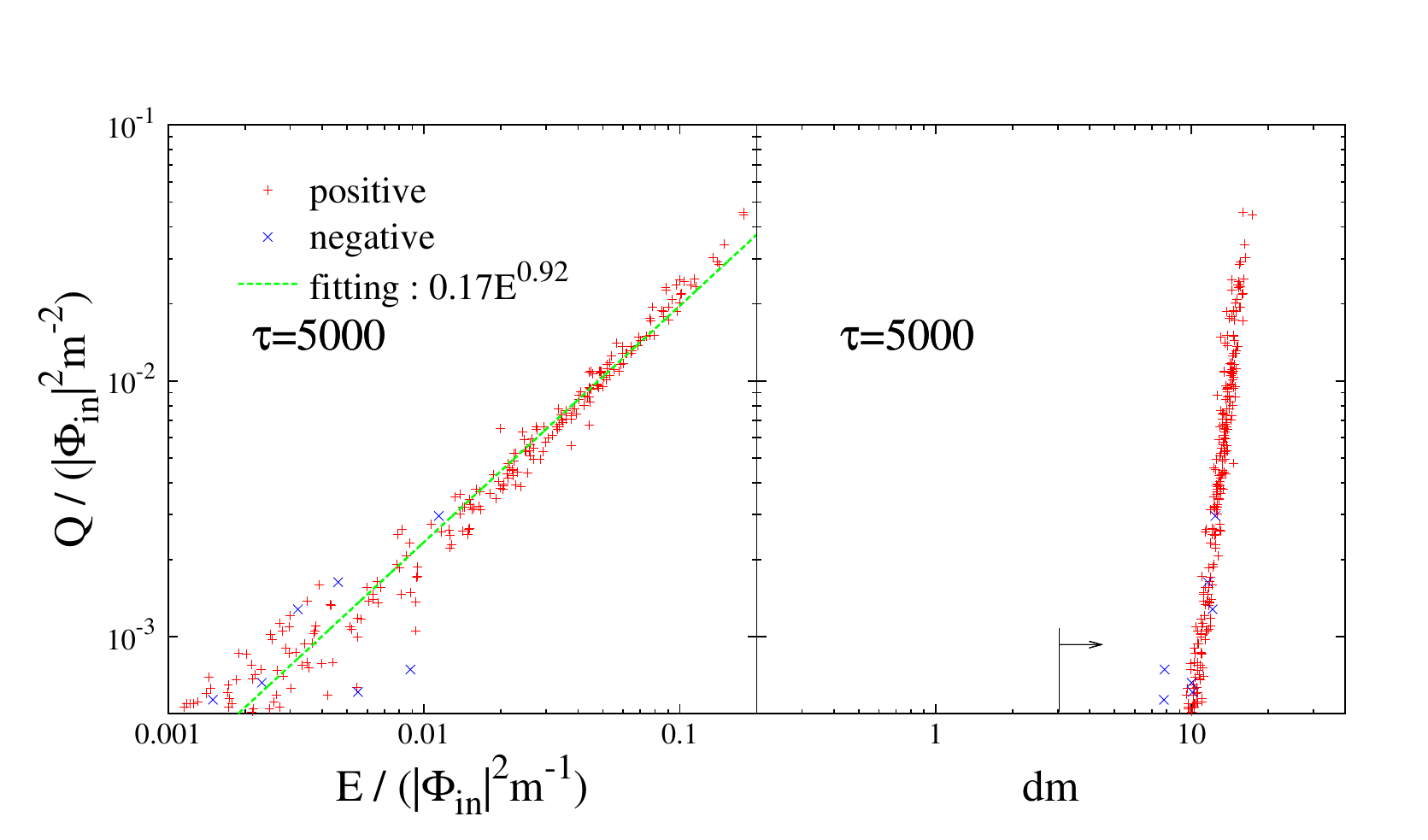}
}
\caption{The charge and the size (diameter $d=2R$) of Q-balls with
 parameter set 3D1 and $q_{\rm c}=1.8\times 10^{-7}\unitq$.
 The red `$+$' and the blue `$\times$' represent positive Q-balls and
 negative ones at $\tau=5000$, respectively.
 The green dashed line represents the fitting formula,
 $|Q|/(\unitQQQ) = 0.17[E/(\unitEEE)]^{0.92}$.
 The arrow standing on the horizontal axis represents the grid size at
 $\tau=5000$ ($\Delta x = 2.99 m^{-1}$).
} 
 \label{fig:qes3D}
\end{figure}

Following the previous discussion in Sec.~\ref{subsubsec:qes2D}, we
investigate the relations among the charge and energy, and size.
The left panel of Fig.~\ref{fig:qes3D} shows the charge and the energy
of the Q-balls in the 3D1 simulation. Also in this 3D system, we can confirm
that these quantities are strongly correlated, having an almost linear
relation, $|Q|/(\unitQQQ)\simeq 0.17[E/(\unitEEE)]^{0.92}$
for $|Q|>10^{-3}\unitQQQ$, as expected. 
Note however that the proportional coefficient is smaller than the one in
the 2D cases. This small coefficient arises from the fact that the
cosmic friction term in Eq.~(\ref{eq:ADevo}) is given by the spatial
dimension. Thus the AD field tends to have a larger ellipticity in the
3D case compared to the 2D case, for a given initial condition in the
phase space. As pointed out in Sec.~\ref{subsubsec:qes2D},
these objects we identified as Q-balls may be slightly excited states of
the Q-balls, and it is expected that those objects will gradually
approach the Q-ball solution in a much longer time scale than
the current limitation of our numerical analysis.

In the right panel of Fig.~\ref{fig:qes3D}, we plot the relation 
between the size (diameter $d=2R$) and the charge (for the definition of
the diameter, see the second paragraph in Sec.~\ref{subsubsec:qes2D}).
The Q-ball solution
has a certain size determined from the instability band shown in
Eq.~(\ref{eq:inst}). We have found that all the Q-balls at least above
$Q>10^{-3}\unitQQQ$ have an almost same size, $d=2R\sim 10m^{-1}$, and
that the size is insensitive to the dimension since this is almost same
as the one in the 2D system.
The spatial resolution in this simulation is presented as an
arrow standing on the horizontal axis of Fig.~\ref{fig:qes3D}, being
much smaller than the typical size of a Q-ball. Hence the Q-balls can be
successfully identified on lattices also in the 3D system.

\subsubsection{K-dependence}
\label{subsubsec:K3D}

\begin{figure}[!ht]
\centering{
\includegraphics[bb=0 0 283 283,width=8cm]{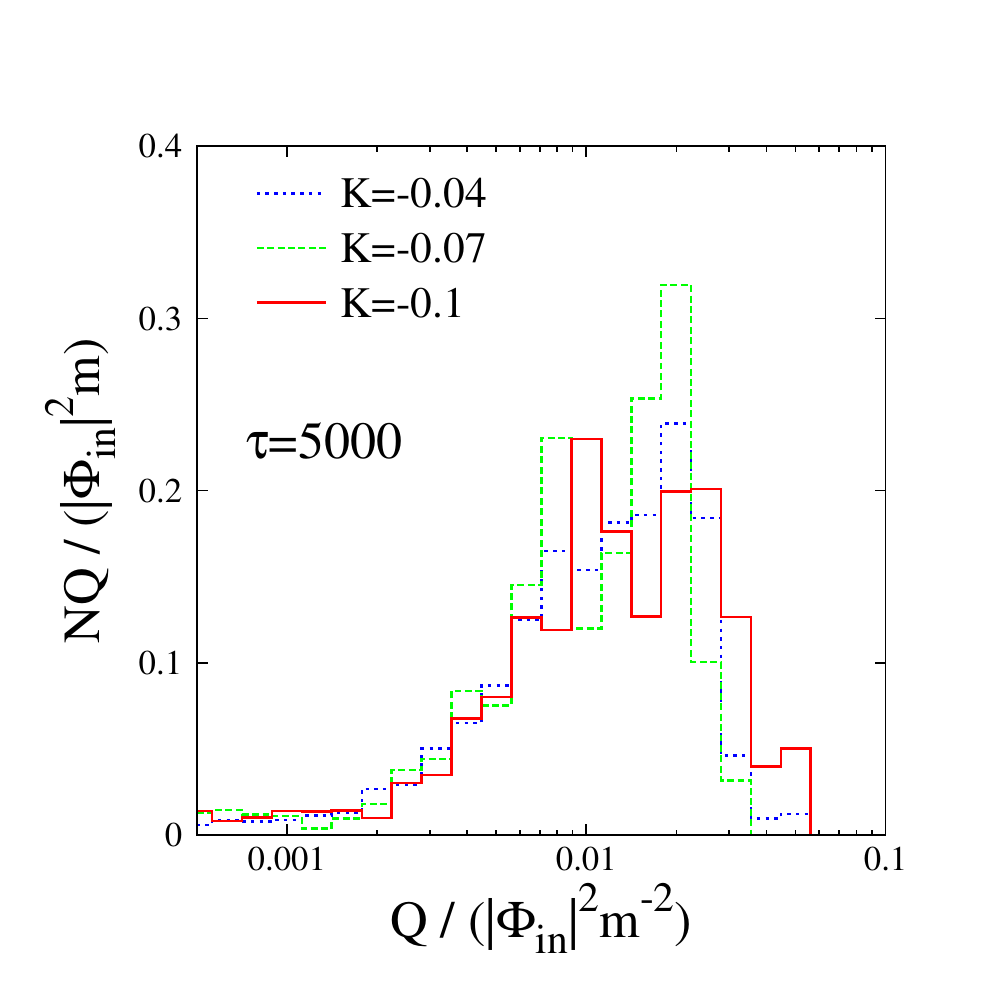}
}
\caption{Charge distributions of Q-balls with
$K=-0.1$ [3D1]$, q_{\rm c}=1.8\times 10^{-7}\unitq$ (red solid),
$K=-0.07$ [3D2]$, q_{\rm c}=1.2\times 10^{-7}\unitq$ (green dashed), and
$K=-0.04$ [3D3]$, q_{\rm c}=6\times 10^{-8}\unitq$ (blue dotted).
}
 \label{fig:Kcharge3D}
\end{figure}

The authors of Ref.~\cite{Kasuya:2001hg} have claimed that the
largest charge 
of Q-balls in the 3D system depends on $K$ as 
$Q^{\rm (KK)}_{\max}\propto |K|^{-1/2}$. To investigate this, we plot
the charge distributions with $K=-0.1, -0.07$ and $-0.04$ in
Fig.~\ref{fig:Kcharge3D}. These parameter sets are shown as 3D1, 3D2 and
3D3 in Table \ref{tab:3D}. Note that, in case with $K=-0.04$, the size
of Q-ball becomes also larger. Hence we take a larger box than the other
cases.

From Fig.~\ref{fig:Kcharge3D}, we cannot see any significant difference
among different values of $K$. 
Suffice it to say that the distribution for $K=-0.04$ might be
marginally skewed to large charges. However, to make this point clear,
more realizations are required. If this is true, the deviation from 
the result of Ref.~\cite{Kasuya:2000wx} may rather widen because their result,
$Q^{(KK)}_{\rm max}=0.012 \unitQQQ$ for $K=-0.01$, would be smaller in the
case with $K=-0.1$.

\begin{figure}[!ht]
\centering{
\includegraphics[bb=0 0 283 283,width=8cm]{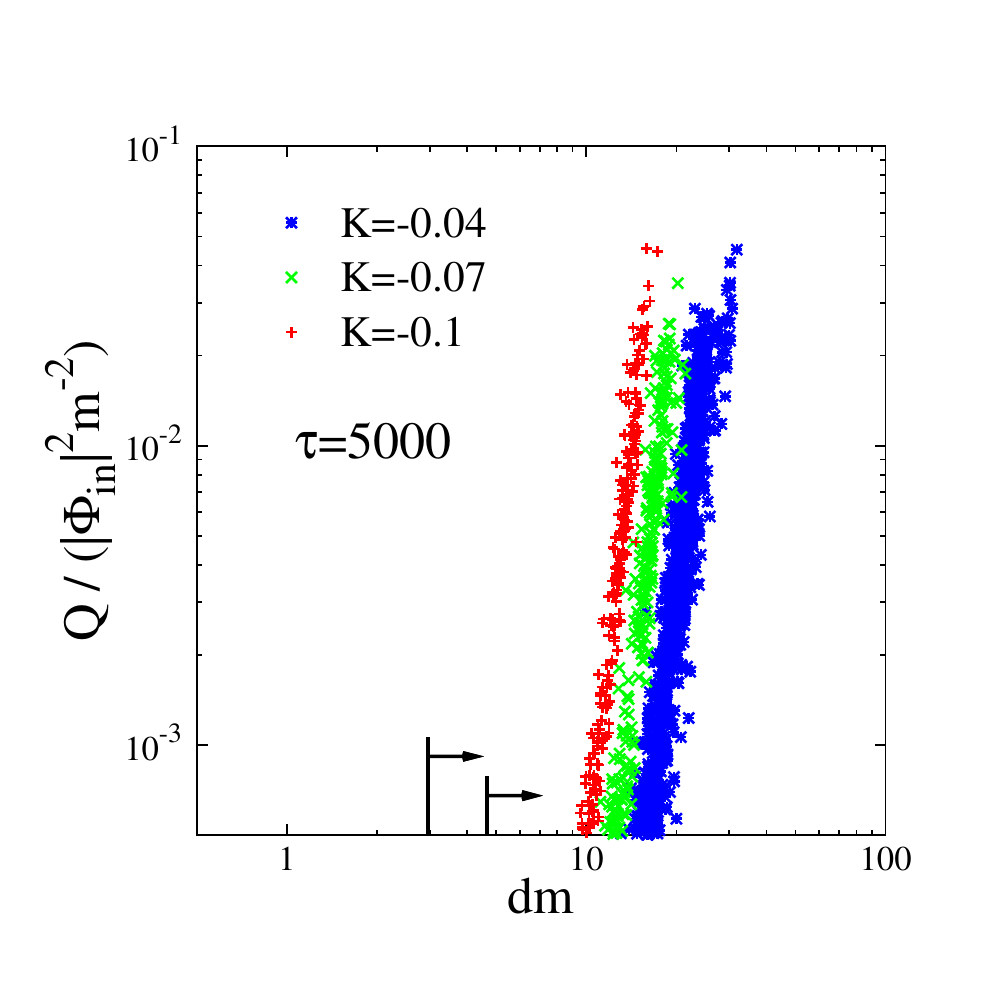}
}
\caption{The charge and the size (diameter $d=2R$) of Q-balls with
$K=-0.1$ [3D1]$, q_{\rm c}=1.8\times 10^{-7}\unitq$ (red `$+$'),
$K=-0.07$ [3D2]$, q_{\rm c}=1.2\times 10^{-7}\unitq$ (green `$\times$'), and
$K=-0.04$ [3D3]$, q_{\rm c}=6\times 10^{-8}\unitq$ (blue `$*$'). 
The arrows standing on the horizontal axis represent the grid size at
 $\tau=5000$ for $K=-0.1, 0.07$ simulations (tall one,
 $\Delta x = 2.99 m^{-1}$) and for $K=-0.04$ simulation (short one,
 $\Delta x = 4.79 m^{-1}$).
}
 \label{fig:Kqs3D}
\end{figure}

The size (diameter $d=2R$) and the charge of Q-balls in simulations
3D1-3D3 are shown in Fig.~\ref{fig:Kqs3D}.
Focusing on the large Q-balls with $Q=Q^{\rm 3D}_{\rm peak} \pm 10\%$ in each
result, we can read the typical sizes as 
$d=2R\sim 15m^{-1}, 19m^{-1}, 25m^{-1}$ for $K=-0.1, -0.07, -0.04$, 
respectively. Thus we confirmed that the size approximately scales as
$R\sim |K|^{-1/2}m^{-1}$. More specifically, we obtain 
$R \sim 2.4|K|^{-1/2}m^{-1}$ at $\tau=5000$. 

\subsubsection{Initial elliptic orbit}
\label{subsubsec:elliptic3D}

As discussed in Sec.~\ref{subsubsec:elliptic2D}, for $\epsilon \ll 1$,
the initial orbit of the AD field becomes elliptic in the phase space.
The resultant extreme energy-to-charge ratio 
$E/mQ \sim 1/\epsilon \gg1$
becomes smaller as a result that the Q-balls discard the
excessive energy. In this process, the negative Q-balls are produced to
conserve the total charge.

\begin{figure}[!ht]
\begin{tabular}{cc}
\begin{minipage}{5.5cm}
\includegraphics[bb=0 0 640 640,width=5.5cm]{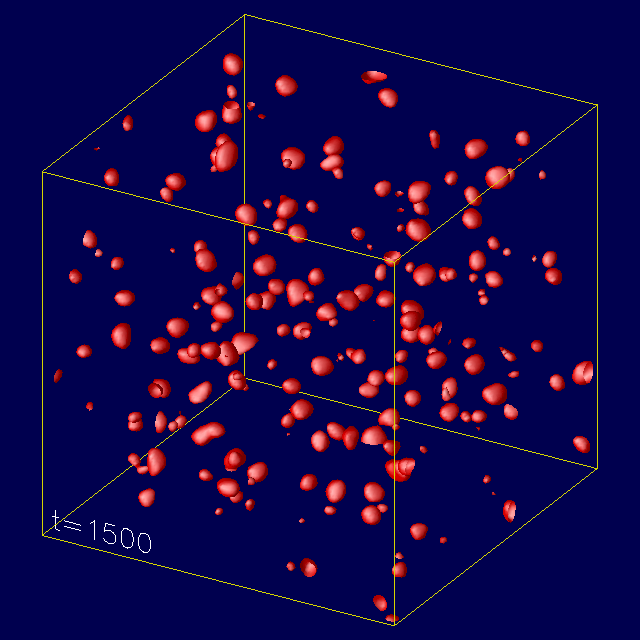}
(a) $\tau=1500$
\end{minipage}
\begin{minipage}{5.5cm}
\includegraphics[bb=0 0 640 640,width=5.5cm]{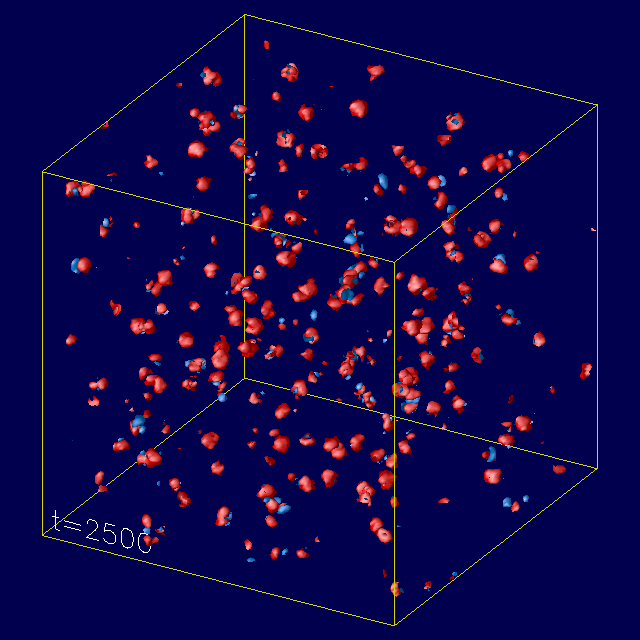}
(b) $\tau=2500$
\end{minipage}
\begin{minipage}{5.5cm}
\includegraphics[bb=0 0 640 640,width=5.5cm]{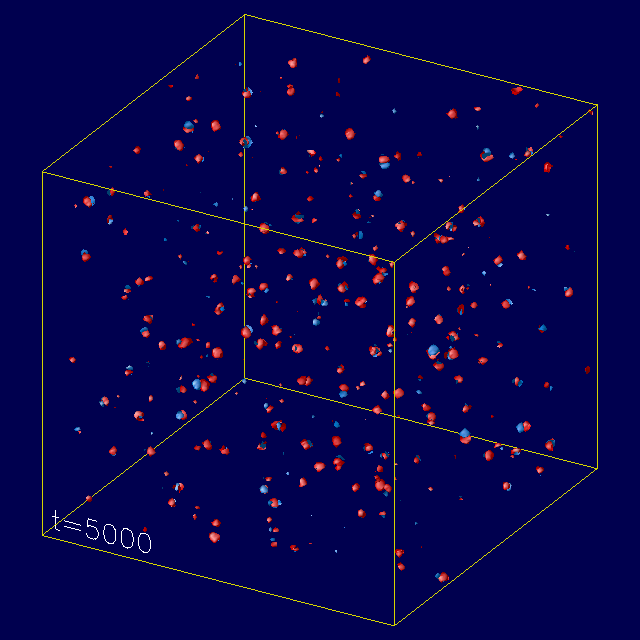}
(c) $\tau=5000$
\end{minipage}
\end{tabular}
\caption{The isodensity surface of the charge density with 
 $q= \pm10^{-6}\unitq$ in the case of $\epsilon=0.1$. 
 Note that the absolute value of this criterion is 60 times
 larger than $q_c$ used in the following analysis for $\epsilon=0.1$
 in order to make the plots clear. The red represents the positive Q-balls,
 and the cyan the negative ones. For these visualizations, we used OpenDX.
} 
 \label{fig:structure3De01}
\end{figure}

\begin{figure}[!ht]
\begin{tabular}{cc}
\begin{minipage}{5.5cm}
\includegraphics[bb=0 0 640 640,width=5.5cm]{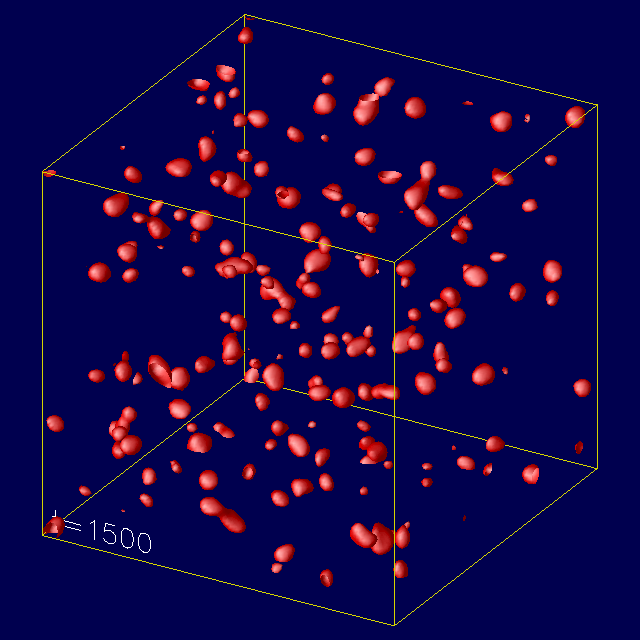}
(a) $\tau=1500$
\end{minipage}
\begin{minipage}{5.5cm}
\includegraphics[bb=0 0 640 640,width=5.5cm]{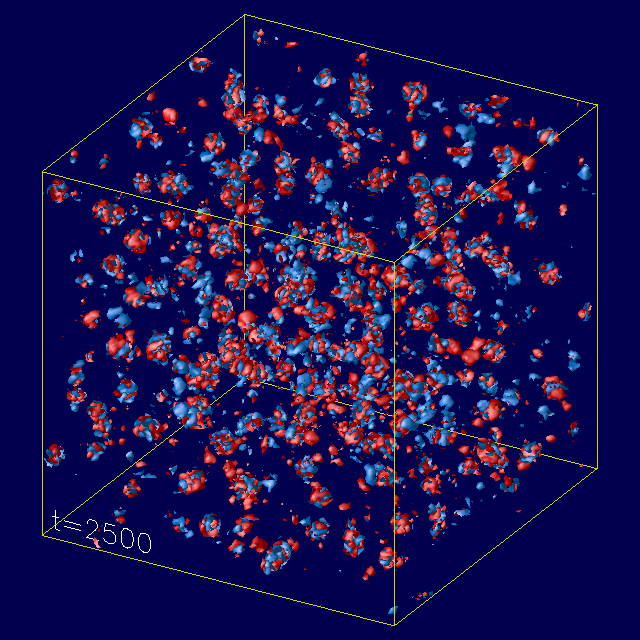}
(b) $\tau=2500$
\end{minipage}
\begin{minipage}{5.5cm}
\includegraphics[bb=0 0 640 640,width=5.5cm]{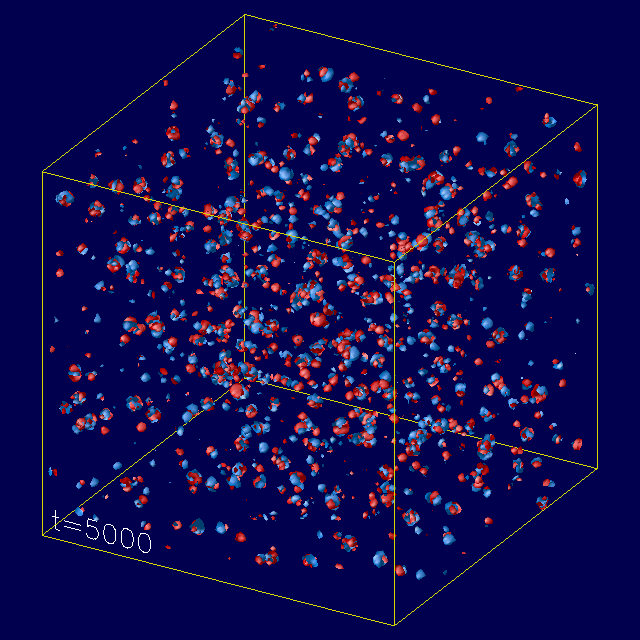}
(c) $\tau=5000$
\end{minipage}
\end{tabular}
\caption{The isodensity surface of the charge density with 
$q=\pm10^{-7}\unitq$ in the case of $\epsilon=0.01$. 
 Note that the absolute value of this criterion is 60 times
 larger than $q_c$ used in the following analysis for $\epsilon=0.01$
 in order to make the plots clear. The red represents the positive Q-balls,
 and the cyan the negative ones. For these visualizations, we used OpenDX.
} 
 \label{fig:structure3De001}
\end{figure}

We have performed numerical simulations with $\epsilon = 0.1$
(3D4) and $\epsilon=0.01$ (3D5). The snapshots of the spatial charge
distribution at $\tau = 1500, 2500$ and $5000$ are shown in
Figs.~\ref{fig:structure3De01} and \ref{fig:structure3De001}, respectively. 
The first formation take place around $\tau\sim 900$, when the filamentary
structure is constructed and torn to small pieces, as in the case
with $\epsilon=1$. Hence we omit the corresponding plots.
The surfaces shown in these figures are the isodensity surfaces with
$q=\pm 10^{-6}\epsilon\unitq$, and the red indicates the positive
Q-balls and the cyan the negative ones. 

In the panel (a), we can see the first-generation Q-balls. As is same as
the 2D cases in Sec.~\ref{subsubsec:elliptic2D}, the all Q-balls have
positive charges. After that, the energy release from the them is
started at $\tau\sim2000$, and as a result, the negative Q-balls appear
around the first-generation Q-balls, which is shown in the
panel (b). This process becomes more violent if $\epsilon$ is smaller.

At $\tau = 5000$ [the panel (c)], the relaxation process has been
almost finished. We can see some isolated negative Q-balls, but most 
of them are still paired with the positive Q-balls. This process to
form the second-generation Q-balls is quite similar to the cases in
the 2D system. Due to the larger spatial dimension, however, it is
observed that a larger number of the negative Q-balls appeared around
one first-generation positive Q-ball. 


\begin{figure}[!ht]
\centering{
  \includegraphics[bb=0 0 510 481,width=13cm]{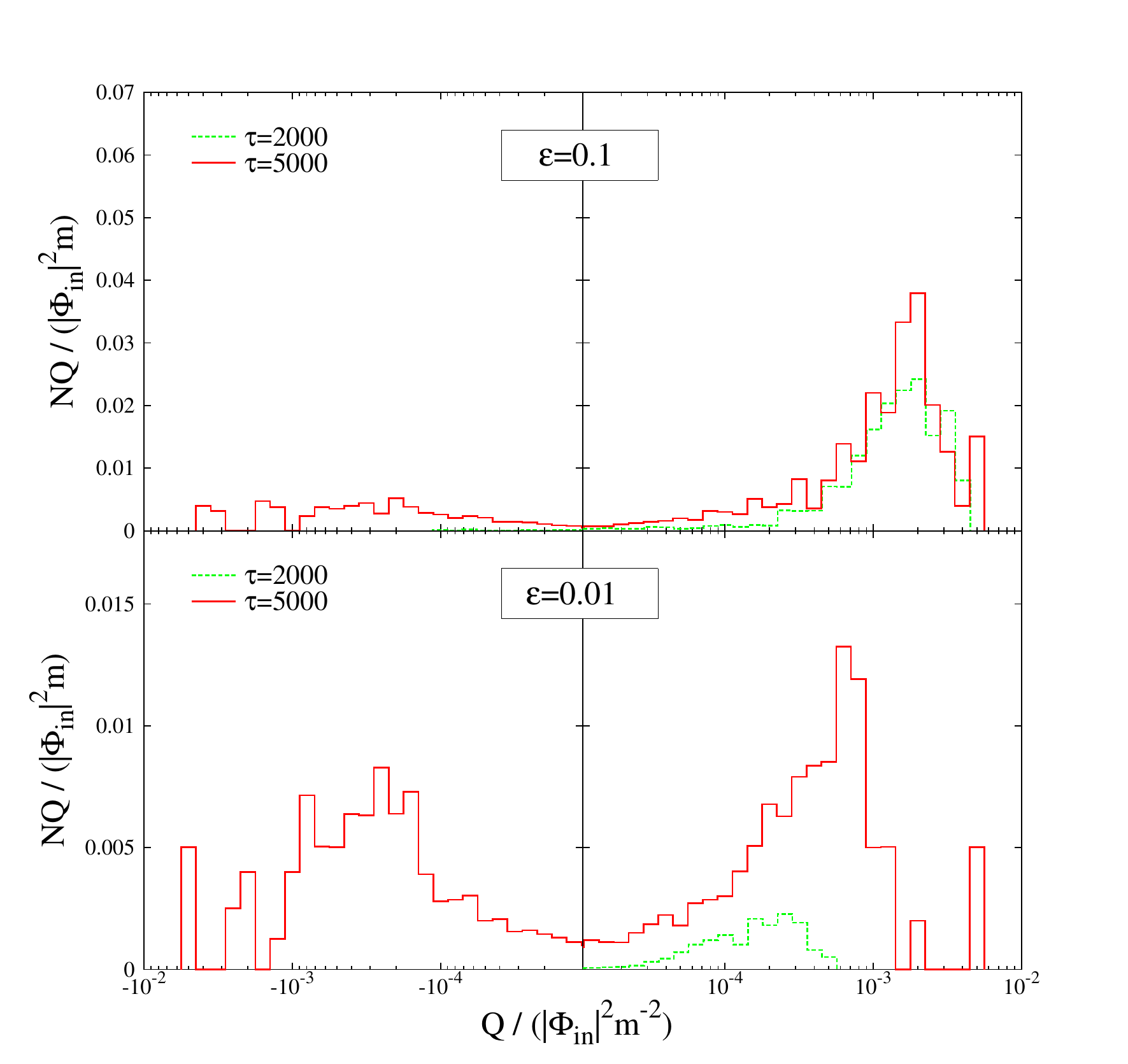}
}
\caption{Charge distributions of Q-balls with 
 $q_{\rm c}=\pm1.8\times 10^{-7}\epsilon\unitq$. The upper panel is the
 case with $\epsilon=0.1$ [3D4], and the lower one is $\epsilon=0.01$ [3D5].
 The left panels are distributions of negative Q-balls and the right
 panels are those of positive Q-balls. 
} 
 \label{fig:eps_charge3D}
\end{figure}

Fig.~\ref{fig:eps_charge3D} shows the charge distributions of the positive and
negative Q-balls. The upper panels are the result with $\epsilon=0.1$, and
the lower ones that with $\epsilon=0.01$. We take the critical charge
$q_{\rm c}=\pm 1.8\times 10^{-7}\epsilon\unitq$. In the case of
$\epsilon=0.1$, almost all the Q-balls are positive, and negative
Q-balls are subdominant. The peak charge is approximately 
$Q^{\rm 3D}_{\rm peak}\sim 2\times 10^{-3}\unitQQQ$ which indicates the scaling
as $Q^{\rm 3D}_{\rm peak}\propto\epsilon$. (Recall that the peak
charge is $Q^{\rm 3D}_{\rm peak}\sim 0.02\unitQQQ$ for $\epsilon = 1$. See
Fig.~\ref{fig:charge3D}.) On the other hand, in the case of
$\epsilon=0.01$, almost same
number of positive and negative Q-balls are produced after 
$\tau \sim 2000$. The peak charge is approximately $2\times 10^{-4}$
until $\tau \lesssim 2000$ which agrees with the scaling. However, the
scaling of the peak charge becomes invalid at $\tau = 5000$. We have
confirmed that it is the Q-balls at the intersections that give dominant
contribution to the peak of the charge distributions. In particular, large
positive and negative Q-balls are found at the intersections in the case
of $\epsilon=0.01$.

\begin{figure}[!ht]
\centering{
\includegraphics[bb=0 0 481 283,width=13cm]{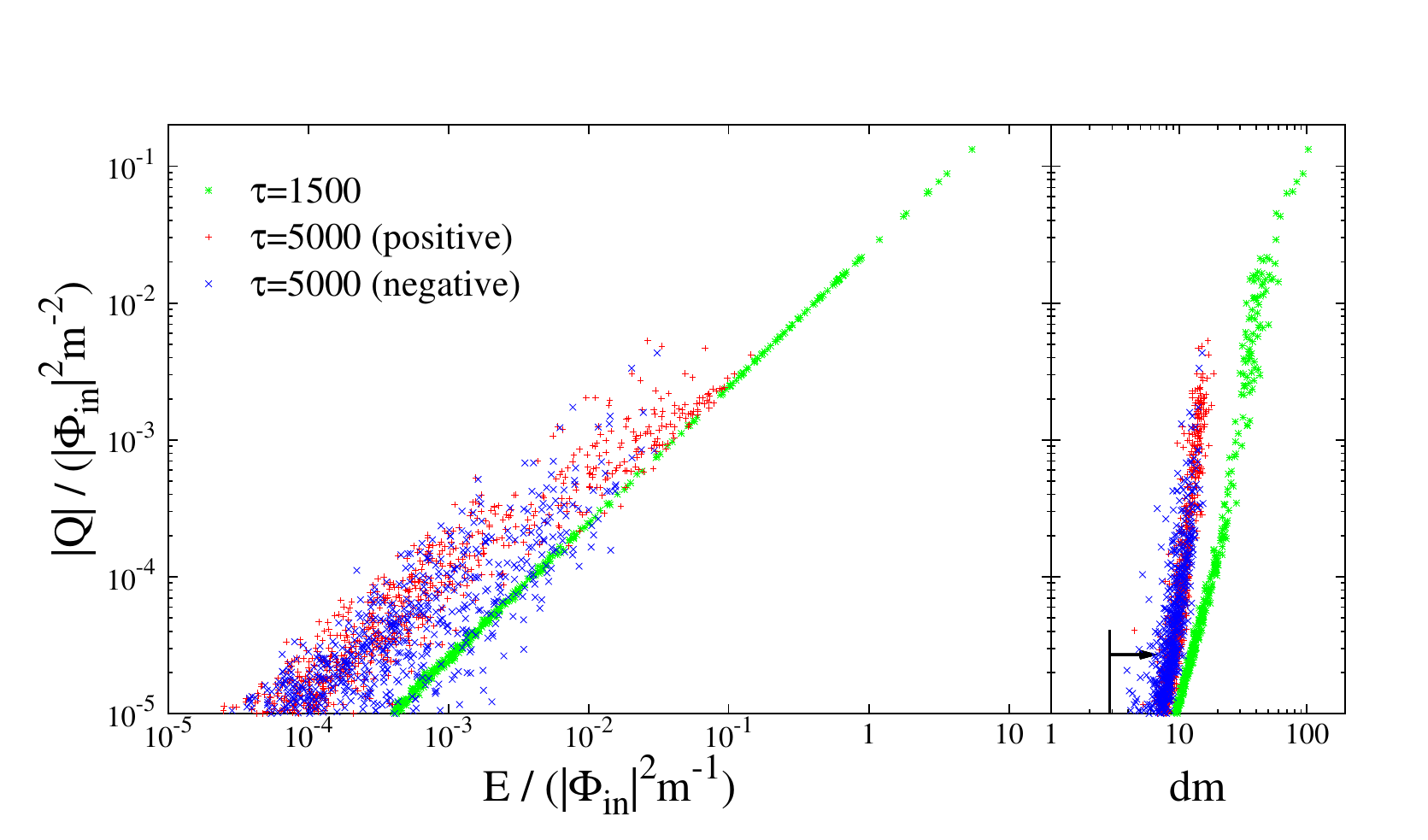}
}
\caption{The charge and the size (diameter $d=2R$) of Q-balls with
 $\epsilon=0.1, q_{\rm c}=\pm1.8\times 10^{-8}\unitq$ [3D4]. 
 The green `*' represents those of positive Q-balls at $\tau=1500$.
 The red `$+$' and the blue `$\times$' represent positive Q-balls and
 negative ones at $\tau=5000$, respectively.
 The arrows standing on the horizontal axis represent the grid size at
 $\tau=5000$ ($\Delta x = 2.99 m^{-1}$).
} 
 \label{fig:eps01_qes3D}
\end{figure}

\begin{figure}[!ht]
\centering{
\includegraphics[bb=0 0 481 283,width=13cm]{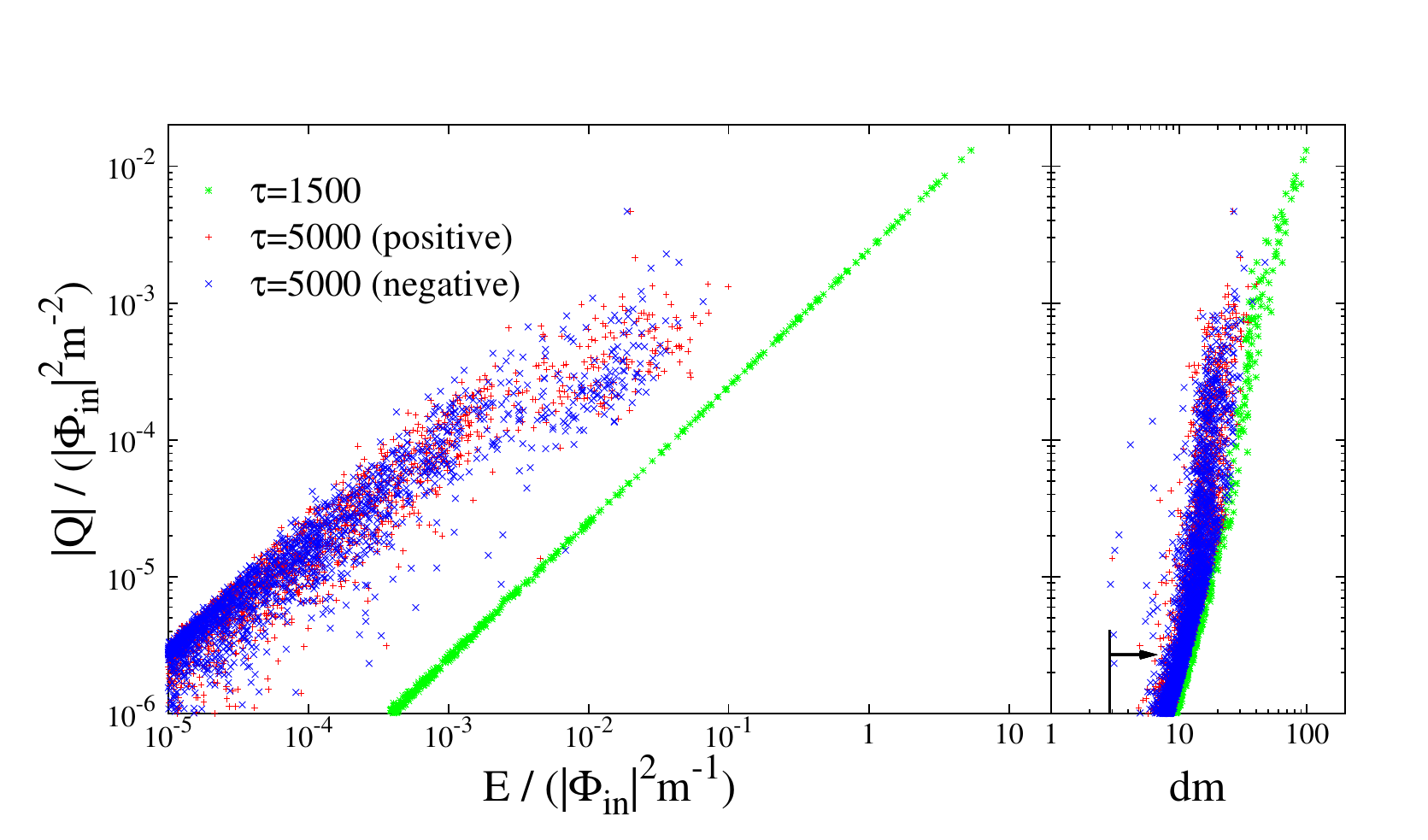}
}
\caption{The charge and the size (diameter $d=2R$) of Q-balls with
 $\epsilon=0.01, q_{\rm c}=1.8\times 10^{-9}\unitq$ [3D5]. 
 The green `*' represents those of positive Q-balls at $\tau=1500$.
 The red `$+$' and the blue `$\times$' represent positive Q-balls and
 negative ones at $\tau=5000$, respectively.
The arrows standing on the horizontal axis represent the grid size at
 $\tau=5000$ ($\Delta x = 2.99 m^{-1}$).
} 
 \label{fig:eps001_qes3D}
\end{figure}

To see the relaxation process where the excited Q-balls change to
the states having lower values of $E/Q$, we show the energy and the
charge of the Q-balls in the 
left panels of Fig.~\ref{fig:eps01_qes3D} and \ref{fig:eps001_qes3D}, and
the relation between the size (diameter) and the charge in the right
panels. In the figures, the green symbols (`*') represent the energy and 
charge of the first-generation (excited) Q-balls at $\tau=1500$, implying
$E/mQ\sim 1/\epsilon$, and 
the red symbols (`$+$') and the blue ones (`$\times$') represent the
positive and the negative Q-balls at $\tau=5000$, respectively.

As is similar to the 2D system discussed in
Sec.~\ref{subsubsec:elliptic2D}, we can see that the energy-to-charge
ratio decreases as time goes, while roughly keeping the proportionality between
$E$ and $Q$, and the sizes become almost universal. 
Looking closely, we found that the minimum energy-to-charge ratio is
approximately $E/mQ\simeq 2$ at $\tau=5000$, and there seems to be a
{\it barrier} which forbids Q-balls to take a smaller value of $E/mQ$.
As is pointed out in Sec.~\ref{subsubsec:elliptic2D}, this ratio may be
determined by the highly non-linear interaction during the secondary
formation process. The relaxation process in which the value of $E/mQ$
eventually becomes $1$ proceeds quite slowly even though there is such
a process. Hence it can be concluded that, in general, 
the second-generation Q-balls acquire values of $E/mQ \sim O(1)$.

To summarize this section, we have observed that the filamentary
structure plays an essential role both in the Q-ball formation and in
the final charge distribution. Large (positive and negative) Q-balls are
created at the intersections of the filaments, while small ones in the sides
and voids.


\section{Applications}
\label{sec:applications}
In this section we show how the Q-ball charge distribution obtained in our numerical simulations affects the
cosmological impact of the Q-balls. We consider the decay and evaporation processes in the following.

\subsection{Fitting formula}
\label{subsec:fitting}
\begin{figure}[!ht]
\centering{
\includegraphics[bb=0 0 283 283,width=8.0cm]{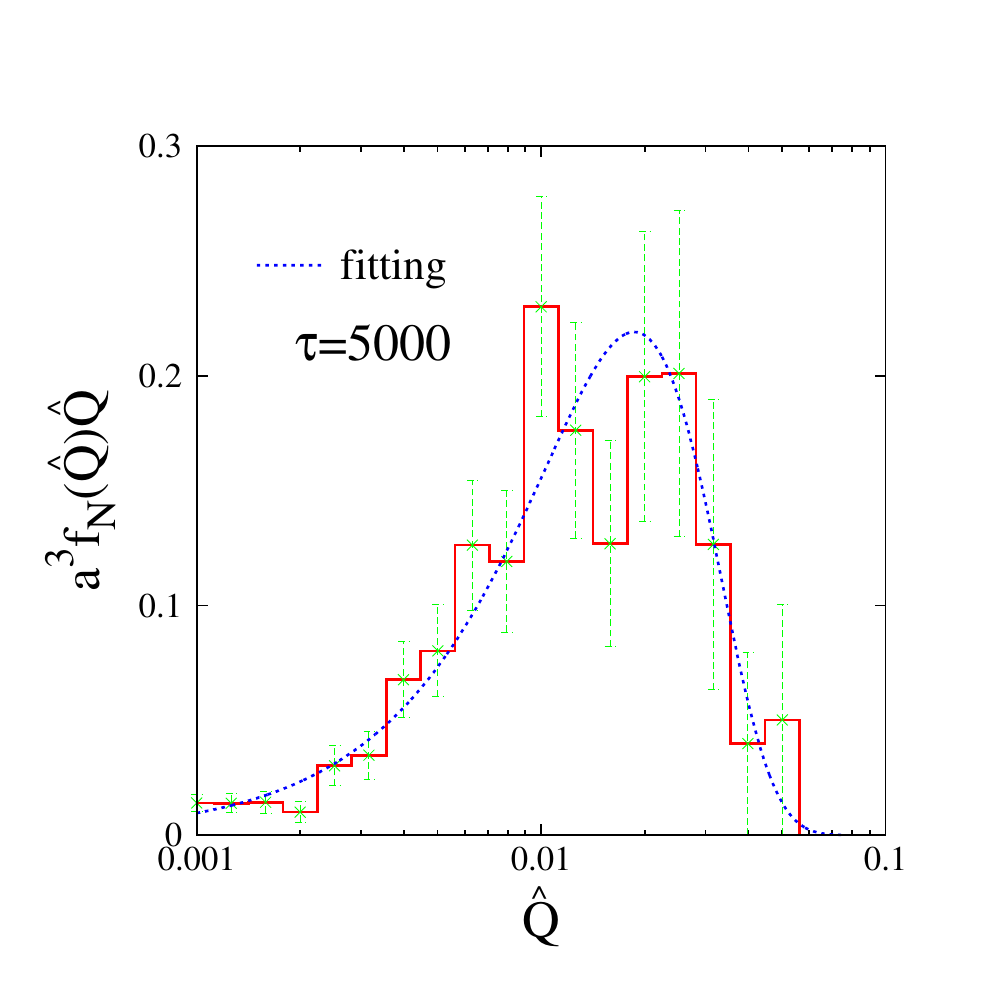}
}
\caption{Charge distribution of Q-balls with parameter set 3D1 and
 $q_{\rm c}=1.8\times 10^{-7}\unitq$ at $\tau=5000$. 
 The red solid is the charge distribution and the green bars represents
 the Poisson error of the number of Q-balls in each bin, namely, $\pm\sqrt{N}$.
 The blue dotted line is the fitting function given as
 $a^3f_{Q}(\hat{Q})\hat{Q}$ where $f_N(t,Q)$ is defined in
 Eq.~(\ref{eq:fit}), and $\hat{Q}$ denotes
 ${\hat Q} \equiv Q / (|\Phi_{\rm in}|^2/m^2)$.
}
 \label{fig:fitting3D}
\end{figure}

First let us give a fitting formula of the charge distribution. For
simplicity we hereafter focus on the case of $\epsilon = 1$. The
distribution function $f_N(t,Q)$ is defined such that the number density
of Q-balls with a charge between $Q$ and $Q+\delta Q$ at the time $t$ is
equal to $f_N(t,Q) \delta Q$. In terms of $N(Q)$, the number of Q-balls
with a charge $Q$ in a comoving volume, it is written as 
%
\begin{equation}
f_N(t,Q) \;=\; a(t)^{-3} N(Q). 
\end{equation}
%
Recall that the scale factor is normalized as $a(t_{\rm in}) = 1$, where
$t_{\rm in} = 2/(3m)$. The numerical result of $N(Q)Q$ is shown in
Fig.~\ref{fig:charge3D}. In order to obtain a fitting formula, we adopt
the following form, 
%
\begin{equation}
a(t)^3 f_N(t,Q) \;=\; N(Q) = x_1 m^3 {\hat Q}^{x_2-1} \exp{[-x_3\, {\hat
 Q}^2]}, 
\label{eq:fit}
\end{equation}
%
where ${\hat Q} \equiv Q / (|\Phi_{\rm in}|^2/m^2)$. Assuming that the
dominant source for the statistical error comes from $N$ following the
Poisson distribution, we have fitted the above form to the charge
distribution at $\tau = 5000$ shown in Fig.~\ref{fig:charge3D}, and
obtained obtained $x_1 \simeq 71.2$, $x_2 \simeq 1.29$, and
$x_3 \simeq 1.86 \times 10^3$. The fitting function and the charge
distribution are shown in Fig.~\ref{fig:fitting3D} where the vertical
axis is $a(t)^3f_{N}(t,Q)Q$ at $\tau=5000$, being equivalent to
Fig.~\ref{fig:charge3D}. Note that the above fitting formula is valid
only for $\epsilon = 1$ and the charge $Q$ between 
$0.001 \lesssim {\hat Q} \lesssim 0.1$. According to the fitting 
formula, the charge distribution $a(t)^3f_{N}(t,Q)Q$ at $\tau=5000$ is
peaked at
%
\begin{equation}
Q_{\rm peak}^{\rm 3D, \epsilon = 1}\;\simeq\;1.9 \times 10^{-2} \lrfp{|\Phi_{\rm in}|}{m}{2},
\end{equation}
%
which is greater than the result of Ref.~\cite{Kasuya:2000wx} by about $60\%$.

\subsection{Decay}
\label{subsec:decay}
The decay of the Q-balls proceeds only in the vicinity of the surfaces,
and the decay rate is known to be bounded above~\cite{Cohen:1986ct},
%
\begin{equation}
\Gamma_Q\;\equiv\; \frac{1}{Q} \left|\frac{dQ}{dt}\right| \leq \frac{\omega^3 A}{192 \pi^2}\,,
\end{equation}
%
where $A$ is the surface area and $\omega$ denotes the energy per unit
charge of the Q-ball, respectively. According to
Ref.~\cite{Cohen:1986ct}, the decay rate is indeed almost saturated. In
the case of the gravity-mediation type Q-ball, the decay rate is then
given by 
%
\begin{equation}
\Gamma_Q \;\simeq\; \frac{m}{24 \pi |K| Q},
\end{equation}
%
where we have used $\omega \simeq m$ and $A = 4 \pi R_Q^2 \simeq
8\pi/(|K| m^2)$. Note that the decay rate is a decreasing function of
$Q$; larger Q-balls are more long-lived. 

Now let us estimate the Q-ball contribution to the total energy at the decay. 
Those Q-balls between $Q$ and $Q+\delta Q$ decay when $H=\Gamma_Q(Q)$,
emitting the radiation density $\delta \rho_Q(Q)$: 
%
\begin{equation}
\frac{\delta \rho_Q}{\rho_{\rm total}} \;=\; \frac{m Q \,f_N(\Gamma_Q^{-1},Q) \delta Q}{3 \Gamma_Q^2 M_P^2},
\end{equation}
%
where $\rho_{\rm total}$ denotes the total energy density at the
decay. Using (\ref{eq:fit}), the distribution at the decay is given by
%
\begin{equation}
f_N(\Gamma_Q^{-1},Q) \;=\; \frac{\Gamma_Q^{\frac{3}{2}}H_R^\frac{1}{2}}{m^2} N(Q),
\end{equation}
%
where $H_R$ denotes the Hubble parameter at the reheating. Thus the
fractional contribution is proportional to $Q^{3/2} N(Q)$, which is
multiplied by an additional factor $Q^{1/2}$ with the charge
distribution. Although the peak may be shifted due to the additional
factor $Q^{1/2}$, we may neglect it in practice. We may use the charge
at the peak to estimate the decay rate of the Q-ball. The decay rate
then turned out to be about $60\%$ smaller than the existing
result~\cite{Kasuya:2000wx}. 

\subsection{Evaporation}
\label{subsec:evaporation}

Through the interaction with the ambient thermal plasma, some amount of
charges are evaporated from
Q-balls~\cite{Laine:1998rg,Banerjee:2000mb}. For a given thermal
history, we can estimate the evaporated charge $\Delta Q$. For the
gravity-mediation type Q-balls, the evaporated charge is given by 
%
\begin{equation}
\Delta Q \sim 6 \times 10^{17} \lrfp{|K|}{0.1}{-\frac{2}{3}} \lrfp{m}{\rm TeV}{-1},
\label{evapQ}
\end{equation}
%
if the reheating temperature is higher than the weak scale.

Before our work, it was customary to assume that almost all the charge
is absorbed into the largest Q-balls with a charge close to
$Q^{\rm 3D}_{\rm peak}$. Then the fraction of the evaporated charge is
estimated by  $\Delta Q/Q^{\rm 3D}_{\rm peak}$. Since we have obtained a
fitting formula for the charge distribution, $f_N(Q)$ (the time variable 
is dropped for simplicity), we can estimate the fraction more precisely
as 
%
\begin{equation}
\frac{\int {\rm min}\left[Q, \Delta Q \right] f_N(Q) dQ}{\int Q f_N(Q) dQ}.
\end{equation}
%
Using the result (\ref{eq:fit}), the fraction of the evaporated charge
reads $\sim 0.12$ and $0.22$ for $\Delta Q/Q^{\rm 3D}_{\rm peak} = 0.1$
and $0.2$, respectively. Thus the evaporated charge turns out to be
$O(10) \%$ larger than the previous result.\footnote{Note that we here
focus on the difference arising only from the crude approximation of
neglecting the charge distribution. If we take account of the difference
in $Q^{\rm 3D}_{\rm peak}$, the discrepancy becomes larger, but in the
opposite direction.} Since the fitting formula may underestimate the
charge distribution for $Q < 0.1 Q^{\rm 3D}_{\rm peak}$, the discrepancy
may be actually much greater for $\Delta Q/Q^{\rm 3D}_{\rm peak} \ll
0.1$. If this is the case, it may be an obstacle for realizing a large
hierarchy between the baryon and lepton asymmetries through the
Q-balls~\cite{Kawasaki:2002hq}. 

\section{Conclusion}
\label{sec:conclusion}

In this paper, we have studied physical properties of the
gravity-mediation type Q-balls in the AD mechanism on 1D, 2D and 3D
lattice simulations {\it with} the cosmic expansion. 

We have found that including cosmic expansion has a crucial effect on
the Q-ball formation and its subsequent evolution, using the 1D lattice
simulations. In the non-expanding background, the Q-ball interactions
through collisions never decouple until one large Q-ball is formed.  In
fact, we confirmed the finite box effect in our simulations without
cosmic expansion. On the other hand, in the expanding background, the
major collisions among large Q-balls do not occur frequently. Even in
the cases of the 2D and 3D simulations, we expect that, without the
cosmic expansion, the Q-ball evolution would be quite different. 

In various setups with the cosmic expansion, we have performed
simulations in 2D and 3D systems to obtain the charge distributions of
the Q-balls. For the initial circular orbit with $\epsilon = 1$, the
charge at the peak of the distribution is $Q^{2D}_{\rm peak}\simeq0.1\unitQQ$
and $Q^{3D}_{\rm peak}\simeq1.9\times 10^{-2}\unitQQQ$.

We have found that filamentary structure is formed when the initial
fluctuation becomes of order unity, as pointed out in
Ref.~\cite{Enqvist:2000cq}. One of our new observations is that this
filamentary structure plays an important role in the Q-ball formation
and affects the final charge distribution. In particular, the Q-balls
forming the peak of the distribution are mainly produced at the intersections
of the filaments, while Q-balls produced from the side and void region
have generically small charges and do not give main contribution to the
charge distribution. Note however that these small Q-balls overwhelm the
large Q-balls  in the number. This argument holds true for a small value
of $\epsilon$, in which large positive and negative Q-balls are mostly
produced from the excited Q-balls at the intersections. (We call the latter
as the first generation Q-balls and the former as the second generation
Q-balls in the text.) 

We have compared the empirical formula derived in
Ref.~\cite{Enqvist:2000cq} to our numerical results on the charge (not
number) distribution  on the 2D lattice and found that, if we fit the
formula at the peak of the charge distribution, it underestimates the
charge distribution at smaller $Q$.  In contrast to the claim made by
Ref.~\cite{Enqvist:2000cq}, therefore, we conclude that these small
Q-balls do not play any significant role to determine the Q-ball
distribution. We also mention here that we stopped following the
evolution once the Q-balls become no longer resolved compared with our
simulation grid size, in order to avoid mis-identification of background
fluctuations with the Q-balls as well as unphysical effects on the
Q-ball evolution. This also may be the reason for the difference of ours
from the results in the past.  

Our findings, especially detailed charge distribution of the Q-balls,
will be very important for the AD baryogenesis and related topics,
because the Q-balls are an essential ingredient of the mechanism. For
instance, the Q-ball decay temperature is modified accordingly, which
affects the resultant dark matter density in a scenario that the
lightest SUSY particles are produced from the Q-ball decay. Also the
charge distribution gives a lower bound on the evaporated charge, which
in turn makes it difficult to realize a large hierarchy between the
baryon and lepton asymmetries in the scenario using the L-balls.  

Lastly we mention the limitation of our numerical simulations. Strictly
speaking,  the Q-balls we identified in the simulations are in a excited
state, in a sense that the $E/mQ$ ratio is slightly larger than $1$. We
expect that these Q-balls will in the end relax to the Q-ball solution,
emitting the excessive energy. Also we have neglected the interactions
of the AD field with gauge fields and other matter fields. We leave
these issues as well as further exploration of the application of our
results for future work. 


\begin{acknowledgments}
We thank S. Kasuya for helpful discussions. The work of 
F.T. was supported by JSPS Grant-in-Aid for Young Scientists (B)
(21740160). M.K. and F.T. were supported by the Grant-in-Aid for
 Scientific Research on Innovative Areas (No. 21111006). 
This work was supported by 
World Premier International Center Initiative (WPI Program), MEXT, Japan.
\end{acknowledgments}

\appendix

\section{Numerical schemes}
\label{appendix:numerical}

The Q-balls are formed through the dynamical instability of the tiny
fluctuations induced at the initial time. Until the formation epoch of
the Q-balls, it takes a long time, typically a few hundreds of
oscillation periods of the background field. Hence a higher-order
numerical scheme is required so that the results does not suffer from
the global error when we follow the evolution of the AD field.
In this paper, we used a 6th-order symplectic integrator developed by
Yoshida \cite{Yoshida:1990zz}. 

If we write the time evolution for a time step $\Delta t$ as a mapping,
$\Phi(t+\Delta t, \xx) = S(\Delta t)\Phi(t, \xx)$, the well-known 2nd-order
leap-frog scheme can be described as 
%
\begin{equation}
  S_{\rm 2nd}(\Delta t) = e^{\frac{1}{2}\Delta t A}e^{\Delta t B}e^{\frac{1}{2}\Delta t A},
\end{equation}
%
where $\Delta t$ is the time step, and the operators $e^{c\Delta t A}$ and
$e^{c\Delta t B}$ map the field and its time-derivative at a time slice $t$
to those at the next time slice $t+c\Delta t$,
%
\begin{align}
 e^{c\Delta t A}  &: \Phi(t, \xx) \to \Phi(t+c\Delta t, \xx) = 
   \Phi(t, \xx) + c\Delta t \dot{\Phi}(t, \xx), \label{eq:eqA} \\
 e^{c\Delta t B}  &: \dot{\Phi}(t, \xx) \to \dot{\Phi}(t+c\Delta t, \xx)
 = \dot{\Phi}(t, \xx) + c\Delta t \ddot{\Phi}(t, \xx). \label{eq:eqB}
\end{align}
%
In Eq.~(\ref{eq:eqB}), the second derivative of $\Phi$ is replaced by
$\dot{\Phi}$ and $\Phi$ using the equation of motion of $\Phi$ [see
Eq.~(\ref{eq:ADevo})]. 
Using this symbolic representation, we can describe the 6th-order scheme as
\cite{Yoshida:1990zz}
%
\begin{equation}
  S_{\rm 6th}(\Delta t) = S_{\rm 2nd}(w_3\Delta t)S_{\rm 2nd}(w_2\Delta t)S_{\rm 2nd}(w_1\Delta t)S_{\rm 2nd}(w_0\Delta t)S_{\rm 2nd}(w_1\Delta t)S_{\rm 2nd}(w_2\Delta t)S_{\rm 2nd}(w_3\Delta t),
\end{equation}
%
where the coefficients $w_1, w_2$ and $w_3$ are given in Table
\ref{tab:6thLF} and $w_0 = 1-2(w_1+w_2+w_3)$. As implied in
Eq.~(\ref{eq:eqB}), the function evaluation, $V'(\Phi)$ in the present
case, is included only in $e^{c\Delta t B}$. Therefore this scheme is 7 stages.

%
\begin{table}[!ht]
\begin{tabular}{c|c}
\hline
  $w_1$ & $-1.17767998417887$ \\
  $w_2$ & $0.235573213359357$ \\
  $w_3$ & $0.784513610477560$ \\
\hline
\end{tabular}
\caption{A set of coefficients used in 6th-order symplectic
 scheme \cite{Yoshida:1990zz}.} 
\label{tab:6thLF}
\end{table}
%

As for the spatial derivative, i.e. the Laplacian operator in
Eq.~(\ref{eq:ADevo}), we used the finite difference scheme called as 3-,
5- and 7-point formulas for 1D, 2D and 3D, respectively. For example, in
the 3D cases, we replace the Laplacian by
%
\begin{equation}
 (\nabla^2 \Phi)_{i,j,k} \approx
\frac{1}{h^2}(\Phi_{i-1,j,k} + \Phi_{i+1,j,k} + \Phi_{i,j-1,k} + \Phi_{i,j+1,k} +\Phi_{i,j,k-1} + \Phi_{i,j,k+1} - 6\Phi_{i,j,k}),
\end{equation}
%
where $\Phi_{i,j,k} \equiv \Phi(t, x_i,y_j,z_k)$.  This derivative is
computed at each stage just before the function evaluation.

\section{identification of Q-balls}
\label{appendix:identifier}

In order to identify a Q-ball in the simulation box, we developed an
identification method based on the Marching Cube algorithm
with some modifications. The Marching Cube method has been frequently
used in imaging analyses \cite{Lorensen:1987}. 

We obtain the field data, $\Phi(t,\xx)$, from the simulations, and can
easily compute the charge density, $q(t,\xx)$, on the grid. In this
charge density field, we regard the region in which the charge density
exceeds the critical value, $|q(t, \xx)| > q_c$, as a Q-ball. 
In accordance with the sign, we call positive Q-ball $(q>0)$ and
negative Q-ball ($q<0$). 

\begin{figure}[!ht]
\centering{
\includegraphics[bb=0 0 321 141]{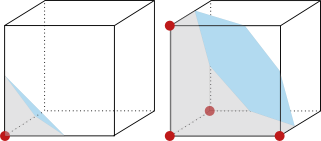}
}
\caption{Two examples of grid cells containing a piece of boundary of a
 Q-ball. The left cell is the case where there is only a grid point
 exceeding the criterion in the cell, represented by red sphere, and the
 right one is a case where there are four points exceeding the
 criterion. The cyan triangle (left) and hexagon (right) are a piece of
 Q-ball boundary, and the gray region represents the inner region of a Q-ball.
} 
 \label{fig:cell}
\end{figure}

Looking closely at a grid cell containing a portion of Q-ball
boundaries, the boundary piece lies in the grid cell as shown, for
example, in Fig.~\ref{fig:cell}. The left cell in the figure is an
example of the case where there is only a grid point exceeding the
critical value, represented by red sphere. The cyan triangle 
is a piece of the boundary of a Q-ball, and the gray region represents
the inner region of the Q-ball. The right cell in the figure is an
example of a case where there are four points exceeding the critical value.
The location of the boundary, namely, the spatial coordinate satisfying
$q(t, \xx) = q_{\rm c}$ on the sides of the grid cell, is 
determined by the linear interpolation between the adjacent grid points.

In accordance with the spatial dimension, there are
several patterns of the shape of Q-ball boundary pieces in a grid cell.
In the 1D case, there are only 1 pattern in which the boundary cuts
a line segment (1D cell). In the 2D cases, we have 4 patterns:
a triangle (or, inversely, a pentagon), a trapezoid, a hexagon, and two
triangles in a 2D cell. Note that the last pattern indicates that there
are parts of two Q-balls in the cell. In the 3D cases, we have 14
patterns including Fig.~\ref{fig:cell}. The all patterns in 3D are shown
in Figure 3 in Ref.~\cite{Lorensen:1987}.

Our algorithm identifying Q-balls is quite simple.
First we seek local peaks of the charge density field satisfying
$|q(t, \xx)| > q_{\rm c}$, which are candidates of
center of individual Q-balls. Then we seek the boundary outward from
the cell containing the peak. At the same time, the energy density and
charge density of each grid cell are computed. Coming to the boundary,
we classify the 
grid cell in the patterns mentioned in the previous paragraph, and
compute the location of the boundary piece. Finally, integrating the
charge density and energy density over the region identified as a
Q-ball, we obtain the charge, energy and the ($D$-dimensional) volume of
the Q-ball. The size (diameter) of the Q-ball is calculated assuming
that the shape of the Q-ball is disk (2D) or ball (3D) with the same
area (2D) or volume (3D). Hence the diameter of the Q-balls is given by
%
\begin{equation}
  d = 
\begin{cases}
  V & \mbox{\rm for 1D},\\
  \sqrt{4V/\pi} & \mbox{\rm for 2D},\\
  \sqrt[3]{6V/\pi} & \mbox{\rm for 3D},
\end{cases} \label{eq:diameter}
\end{equation}
%
where $V$ is the $D$-dimensional volume.

If by chance there are more than two peaks of the charge density field
in the region satisfying $|q(t, \xx)| > q_{\rm c}$, the present scheme
counts redundantly multiple Q-balls. Hence, at the final stage of this
algorithm, the redundant Q-balls are eliminated.

\bibliographystyle{apsrev}

\end{document}